\documentclass[pra,a4paper,twocolumn,showpacs,floatfix,superscriptaddress,amsmath,amsfonts,amssymb,preprintnumbers,citeautoscript,aps]{revtex4}
\pdfoutput=1
\usepackage[T1]{fontenc}\usepackage[latin1]{inputenc}
\usepackage{dcolumn,graphicx,color,booktabs,microtype,afterpage} \graphicspath{{Figures/}}
\usepackage[charter,greekuppercase=italicized]{mathdesign}
\renewcommand{\figurename}{Fig.}
\renewcommand{\tablename}{Table}
\makeatletter\renewcommand{\fnum@figure}[1]{\figurename~\thefigure.}\makeatother
\makeatletter\renewcommand{\fnum@table}[1]{\tablename~\thetable.}\makeatother
\newcount\hh \newcount\mm
\hh=\time \divide\hh by 60
\mm=\hh \multiply\mm by 60 \mm=-\mm
\advance\mm by \time
\def\now{\number\hh:\ifnum\mm<10{}0\fi\number\mm}

\usepackage[colorlinks,plainpages=false,linkcolor=blue,urlcolor=blue,citecolor=blue,pdfpagemode=UseNone,pdfstartview=FitBH]{hyperref}

\newcommand{\cb}{CeB$_\text{6}$}
\newcommand{\clb}{Ce$_{\text{1}-x}$La$_x$B$_\text{6}$}
\newcommand{\cnbx}{Ce$_{\text{1}-x}$Nd$_{x}$B$_\text{6}$}
\newcommand{\cnb}{Ce$_\text{0.7}$Nd$_\text{0.3}$B$_\text{6}$}
\newcommand{\ccb}{Ce$_{\text{1}-x}$La$_x$Cu$_\text{6}$}

\begin{document}

\makeatletter\renewcommand{\ps@plain}{%
\def\@evenhead{\hfill\itshape\rightmark}%
\def\@oddhead{\itshape\leftmark\hfill}%
\renewcommand{\@evenfoot}{\hfill\small{--~\thepage~--}\hfill}%
\renewcommand{\@oddfoot}{\hfill\small{--~\thepage~--}\hfill}%
}\makeatother\pagestyle{plain}

%\preprint{\textit{Preprint: \today, \now. For internal use only, do not distribute.}}%\linenumbers

\title{Doping-induced redistribution of magnetic spectral weight\\ in substituted hexaborides \clb\ and \cnbx}

\author{S.~E.~Nikitin}
\affiliation{Institut f\"ur Festk\"orper- und Materialphysik, Technische Universit\"at Dresden, D-01069 Dresden, Germany}
\affiliation{Max Planck Institute for Chemical Physics of Solids, N\"{o}thnitzer Str. 40, 01187 Dresden, Germany}
\author{P.~Y.~Portnichenko}
\affiliation{Institut f\"ur Festk\"orper- und Materialphysik, Technische Universit\"at Dresden, D-01069 Dresden, Germany}
\author{A.\,V. Dukhnenko}\author{N.~Yu.~Shitsevalova}\author{V.~B. Filipov}
\affiliation{I.\,M.\,Frantsevich Institute for Problems of Material Sciences of NAS, 3 Krzhyzhanovsky str., 03680 Kiev, Ukraine}
\author{Y.~Qiu}
\affiliation{NIST Center for Neutron Research, National Institute of Standards and Technology, Gaithersburg, Maryland 20899, USA}
\author{J.~A.~Rodriguez-Rivera}
\affiliation{NIST Center for Neutron Research, National Institute of Standards and Technology, Gaithersburg, Maryland 20899, USA}
\affiliation{Materials Science and Engineering, University of Maryland, College Park, Maryland 20742, USA}
\author{J.~Ollivier}
\affiliation{Institut Laue-Langevin, 71 avenue des Martyrs, CS 20156, F-38042 Grenoble Cedex 9, France}
\author{D.~S.~Inosov}\email[Corresponding author: ]{Dmytro.Inosov@tu-dresden.de}
\affiliation{Institut f\"ur Festk\"orper- und Materialphysik, Technische Universit\"at Dresden, D-01069 Dresden, Germany}

\begin{abstract}\noindent
We investigate the doping-induced changes in the electronic structure of CeB$_6$ on a series of substituted Ce$_{1-x}R_x$B$_6$ samples ($R$~=~La, Nd) using diffuse neutron scattering. We observe a redistribution of magnetic spectral weight across the Brillouin zone, which we associate with the changes in the Fermi-surface nesting properties related to the modified charge carrier concentration. In particular, a strong diffuse peak at the corner of the Brillouin zone ($R$ point), which coincides with the propagation vector of the elusive antiferroquadrupolar (AFQ) order in CeB$_6$, is rapidly suppressed by both La and Nd doping, like the AFQ order itself. The corresponding spectral weight is transferred to the $X(00\frac{1}{2})$ point, ultimately stabilizing a long-range AFM order at this wave vector at the Nd-rich side of the phase diagram. At an intermediate Nd concentration, a broad diffuse peak with multiple local maxima of intensity is observed around the $X$ point, evidencing itinerant frustration that gives rise to multiple ordered phases for which Ce$_{1-x}$Nd$_x$B$_6$ is known. On the La-rich side of the phase diagram, however, dilution of the magnetic moments prevents the formation of a similar $(00\frac{1}{2})$-type order despite the presence of nesting. Our results demonstrate how diffuse neutron scattering can be used to probe the nesting vectors in complex $f\!$-electron systems directly, without reference to the single-particle band structure, and emphasize the role of Fermi surface geometry in stabilizing magnetic order in rare-earth hexaborides.
\end{abstract}
\pacs{71.27.+a, 75.20.Hr, 75.30.Kz, 75.40.Gb, 78.70.Nx}

\maketitle\enlargethispage{2pt}

\section{Introduction}\vspace{-2pt}

Although knowing the electronic structure plays a crucial role for understanding macroscopic properties of any crystalline material, there are very few experimental techniques that can be used to probe it. Standard macroscopic measurements that are usually used for reconstruction of the Fermi-surface geometry are quantum oscillations (de Haas\,--\,van Alphen and Shubnikov\,--\,de Haas effects), but both of them require very high purity of samples with a long mean free path, which is difficult to achieve for nonstoichiometric chemical compositions, and high magnetic fields, which may significantly change the electronic structure itself in the case of materials with strong spin-orbit coupling or superconductors. In addition, the resulting information is limited to the areas of extremal electron orbits and is therefore insufficient to extract precise values of Fermi momenta. A more direct source of information about Fermi surface geometry is angle-resolved photoelectron spectroscopy (ARPES), which is able to probe the 4-dimensional \mbox{($k_x$, $k_y$, $k_z$ and $\hbar\omega$)} electronic structure, but in turn, this technique is very sensitive to the quality of the sample surface and has poor resolution along the momentum axis perpendicular to the cleavage plane. This limits its applicability to materials with highly 3-dimensional (3D) band structures. Extraction of Fermi-surface nesting vectors from ARPES data is possible and has been successful in many earlier works \cite{StraubFinteis99, SchaeferSing03, InosovZabolotnyy08, TerashimaSekiba09, EvtushinskyInosov10, HeFujita11, InosovEvtushinsky09, KoitzschHeming16}, but it usually relies on a technically demanding fit of the whole low-energy band structure to a tight-binding model with a consequent momentum integration to extract the peaks in a two-particle correlation function. This method is, therefore, indirect.

A handful of earlier works on magnetic heavy-fermion metals, where conduction electrons are involved in the formation of magnetic order, suggest that the low-energy dynamic spin susceptibility $\chi(\omega,\mathbf{Q})$, measured with diffuse neutron scattering, provides direct information about the nesting vectors \cite{InosovEvtushinsky09, KoitzschHeming16, SinhaLander81, AeppliYoshizawa86, StockertFaulhaber04, KadowakiSato04, WiebeJanik07, StockBroholm12, PortnichenkoCameron15, ButchManley15}. Due to the bulk sensitivity of neutron scattering, it therefore serves as a complementary method for probing the 3D electronic structure. In this paper, we apply this method to probe the evolution of the electronic properties of cerium hexaboride, \cb, upon La and Nd substitutions and discuss the changes in its Fermi-surface nesting properties that correlate with the appearance of different ordered phases in the magnetic phase diagram.

Heavy-fermion compounds are a class of correlated $f\!$-electron materials, which continue to attract close attention despite decades of intense investigations due to a number of intriguing physical phenomena, such as unconventional superconductivity \cite{SteglichBredl84, SchlabitzBaumann86, ZhangShelton00book, Pfleiderer09, Stewart17}, quantum criticality \cite{SiSteglich10, StockertSteglich11}, and multipolar ordered phases \cite{SakaiShiina05, NagaoIgarashi06, SantiniCarretta09, PaschenLarrea14, PortnichenkoPaschen16}. A model example of intriguing heavy-fermion physics is \cb, which exhibits a very complex magnetic phase diagram and a rich spectrum of excitations, indicating a delicate balance between different microscopic interactions despite very simple crystal and electronic structures \cite{EffantinRossatMignod85, SluchankoBogach07, CameronFriemel16}. Zero-field phase diagram of \cb\ consists of a complex double-$\mathbf{q}$ antiferromagnetic (AFM) phase below $T_{\text{N}}=2.3$~K \cite{ZaharkoFischer03} and the so-called ``hidden order'' phase II between $T_{\text{N}}$ and $T_{\text{Q}}=3.2$~K, which results from a G-type antiferroquadrupolar (AFQ) ordering of the $\mathcal{O}_{xy}, \mathcal{O}_{yz}$ and $\mathcal{O}_{zx}$ quadrupolar moments of Ce$^{3+}$ ions \cite{Kusunose08, SakaiShiina97, ShiinaShiba97, ShiinaSakai98, SeraKobayashi99}. Most recently, an additional crossover was found in the behavior of the anisotropic magnetoresistance inside phase~II, separating this phase into two regions characterized by opposite signs of the resistance anisotropy with respect to the field direction \cite{DemishevKrasnorussky17}. \nocite{SeraSato87, TayamaSakakibara97, KobayashiSera00, KobayashiYoshino03, YoshinoKobayashi04, MignotRobert09, SchenckGygax07, KunimoriTanida10, FriemelJang15, MatthiasGeballe68, ArkoCrabtree75, BatkoBatkova95} In contrast to the spectrum of a conventional antiferromagnet with Goldstone modes stemming from magnetic Bragg peaks, the excitation spectrum of \cb\ contains several novel features, such as a resonant mode at the $R$ point \cite{FriemelLi12}, which corresponds to the propagation vector of the AFQ phase, and an intense ferromagnetic (FM) excitation with a parabolic dispersion at the $\Gamma$ point \cite{JangFriemel14}.

\begin{figure}[t]
  \includegraphics[width=\linewidth]{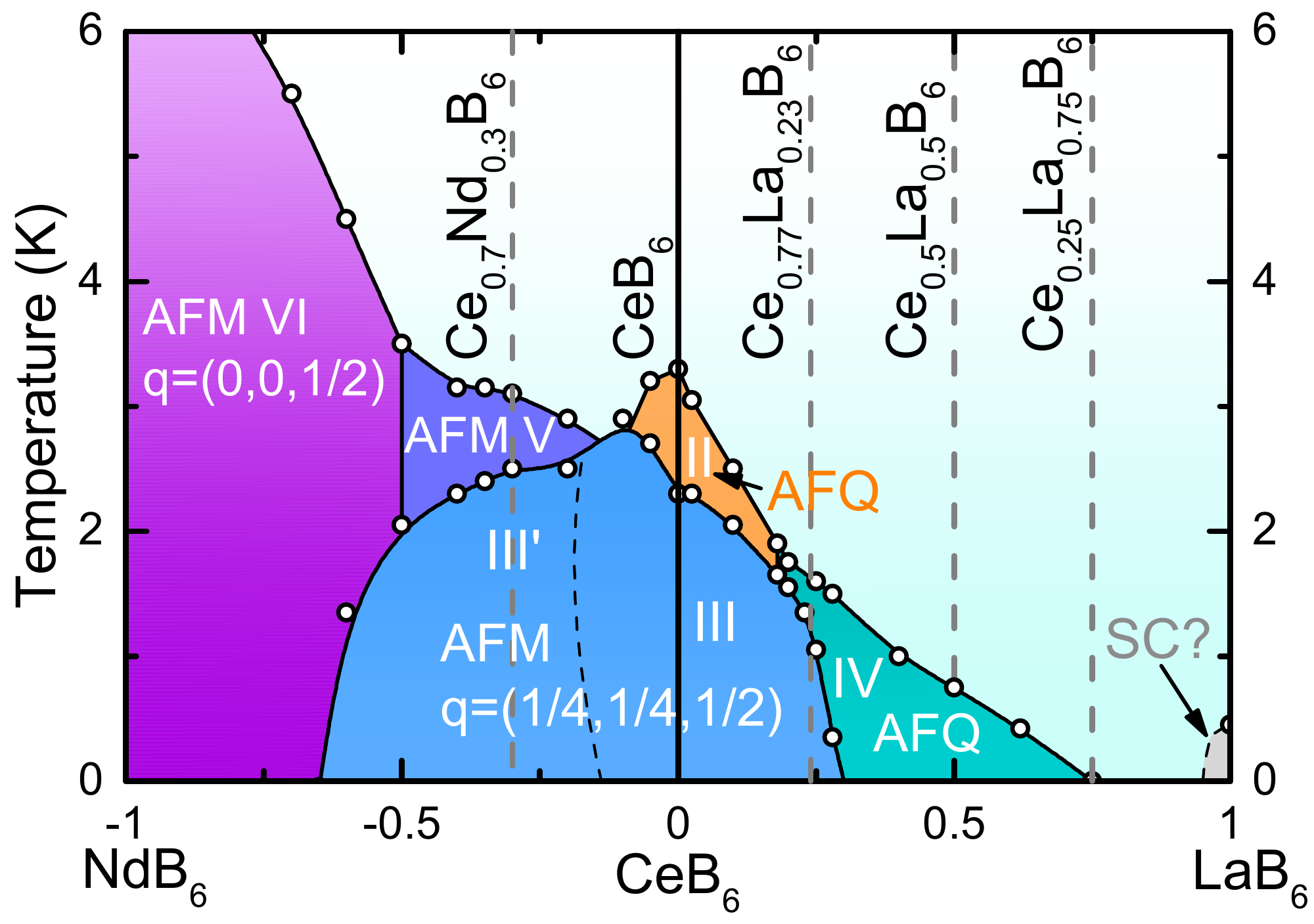}
  \caption{Schematic magnetic phase diagram of the solid solutions Ce$_{1-x}R_x$B$_6$ ($R$~=~La, Nd) after Refs.~\cite{SeraSato87, TayamaSakakibara97, KobayashiSera00, KobayashiYoshino03, YoshinoKobayashi04, SchenckGygax07, KunimoriTanida10, FriemelJang15, MignotRobert09, CameronFriemel16} at zero field. Phases II and IV are associated with the two multipolar phases with AFQ and octupolar ordering. The ``SC'' dome at the bottom-right corner schematically indicates the dubious superconducting phase of the pure LaB$_6$ \cite{MatthiasGeballe68, ArkoCrabtree75, BatkoBatkova95, CameronFriemel16}. Phases III, V and VI are three different types of AFM ordering, realized in \cnbx. Vertical gray lines indicate the sample compositions, measured in our present work.
  }
  \label{PhDsub}
\end{figure}

% Substituted hexaborides
Generally, in heavy-fermion materials, isovalent substitution of magnetic ions with non-magnetic ones leads to a crossover from a coherent Kondo lattice to the dilute Kondo impurity regime, as it was shown for \ccb~\cite{OnukiShimizu85, SumiyamaOda86, KatoSatoh87, SatohFujita89}. Lanthanum doping of \cb\ suppresses both AFM and AFQ phases and induces a new multipolar-ordered phase~IV beyond a quantum critical point (QCP) at $x\approx0.3$, presumably of octupolar character \cite{LoveseyFernandezRodriguez07, KuwaharaIwasa07, KuwaharaIwasa09}. Non-Fermi-liquid behavior has been observed in the vicinity of the QCP \cite{NakamuraEndo06}, and a recent specific-heat investigation of substituted \clb\ emphasizes the importance of multipolar fluctuations that are directly linked to the effective mass of charge carriers \cite{JangPortnichenko17}. However, the nature of the ordered multipoles in phase IV, as well as the underlying interactions \cite{ShibaSakai99, ShibaSakai00, HanzawaTakasaki02, Hanzawa04, SakuraiKuramoto05}, are still a matter of debate.

A substitution of Ce$^{3+}$ with other magnetic rare-earth ions, like Pr$^{3+}$ or Nd$^{3+}$, significantly enriches the phase diagram \cite{KondoTou07, KishimotoKondo05, KobayashiYoshino03, YoshinoKobayashi04, MignotRobert09, MatsumuraKunimori14}. In the simplest case of Nd doping, even a rather low Nd concentration of $\sim$\,0.1 suppresses the AFQ phase at zero field, makes the AFM propagation vector of phase III slightly incommensurate, and creates a new AFM phase~V. At $x\approx0.5$, the order finally changes to conventional AFM stacking of FM layers with the ordering wave vector $\mathbf{q}_0=(00\frac{1}{2})$, like in the pure NdB$_6$ \cite{KobayashiYoshino03, YoshinoKobayashi04, MignotRobert09}. The zero-field magnetic phase diagram of both \clb\ and \cnbx\ is summarized in Fig.~\ref{PhDsub}.

Investigations of the electronic structure of light rare-earth hexaborides have shown that the shape of the Fermi-surface, effective mass of charge carriers, and the number of conduction electrons per unit cell are very similar for both NdB$_6$ and LaB$_6$, due to a strong localization of $4f$ electrons in NdB$_6$ \cite{OnukiUmezawa89, ArkoCrabtree76}. On the other hand, hybridization of Ce $4f^1$ electrons with the conduction band qualitatively modifies the Fermi surface of \cb\ as compared to LaB$_6$~\cite{NeupaneAlidoust15}. Therefore, both La and Nd doping of \cb\ do not simply change the number or magnitude of localized $4f$ magnetic moments, but also induce an effective hole doping, decreasing the number of conduction electrons and modifying the Fermi surface geometry.

In a recent work \cite{KoitzschHeming16}, some of us have investigated electronic structure of pure \cb\ using both ARPES and inelastic neutron scattering (INS). We observed strong spectral intensity at the points, which correspond to the propagation vectors of the AFQ and AFM phases, and additional intensity maxima at the $\Gamma$ and $X$ points. Our analysis of the ARPES data has shown that both order parameters of AFM and AFQ phases are dictated by nesting instabilities of the Fermi surface, and therefore electronic and magnetic structures of \cb\ are closely connected with each other. In this paper, we extend this approach to a series of La and Nd substituted Ce$_{1-x}R_x$B$_6$ solid solutions and use INS to uniformly cover the phase diagram shown in Fig.~\ref{PhDsub} in order to demonstrate that the doping-induced changes in the nesting properties of the Fermi surface correlate with the changes of the ordered phases for compounds with various substitution levels.

\vspace{-2pt}\section{Experimental results}\vspace{-2pt}
\subsection{Details of the experiments}\vspace{-2pt}

In this work we used single crystals of \clb\ ($x = 0, 0.23, 0.5, 0.75$) and \cnb\ with a typical mass of $\sim$\,3\,--\,4~g, specially grown using isotope-enriched $^{11}$B to minimize neutron absorption, as described elsewhere~\cite{FriemelLi12}. The crystal structure is cubic, with a lattice constant of 4.14~\AA\ for pure CeB$_6$, and belongs to the $Pm\overline{3}m$ space group. For thermodynamic measurements of \cnb\ we used small pieces of the same single crystal that was used for INS measurements, cut along $[1\bar{1}0]$. Specific heat and magnetization of \cnb\ were measured using the physical-property measurement system PPMS-6000 and the vibrating-sample magnetometer MPMS3-VSM, respectively.

For neutron scattering measurements, we oriented all samples with their $[1\bar{1}0]$ crystal axes perpendicular to the scattering plane in order to have access to all high-symmetry directions in the $(HHL)$ plane. INS measurements of the parent compound \cb\ were performed using the cold-neutron time-of-flight spectrometer IN5 \cite{OllivierMutka11} at the high-flux research reactor of the Institute Laue-Langevin (ILL) at a temperature of $T=3.2$~K, just above $T_{\text{Q}}$. The incident neutron wavelength was fixed at 5~\smash{\AA} ($E_{\text{i}}=3.27$~meV), which corresponds to an energy resolution (full width at half maximum) of $0.08$~meV at zero energy transfer.

Measurements of substituted samples were performed using the MACS spectrometer at NIST \cite{RodriguezAdler08}. The measurement temperature was fixed just above $T_\text{N}$ for each sample. The MACS spectrometer operates a system of multiple analyzers and detectors that comprise 20 identical channels surrounding the sample. Each channel contains a vertically focusing double-crystal analyzer with two detectors. Such a design implies that one can simultaneously collect data with the given final neutron energy, $E_{\rm f}$, for the spectroscopic channel, and without energy selection for the diffraction channel. This allows us to collect energy-integrated data in parallel to any spectroscopic measurement at no extra cost in acquisition time. In our measurements, we fixed the incident neutron energy $E_{\text{i}}$ to 3.2~meV, which implies an energy resolution of $\Delta{}E=0.15$~meV at the elastic position, and chose an energy transfer of 0.2~meV for the spectroscopic channel to map out the diffuse quasielastic magnetic scattering (QEMS) intensity just above the incoherent elastic line. We used cold Be filters both before and after the sample to suppress higher-order contamination from the monochromator. To distinguish the diffuse magnetic signal from nonmagnetic background scattering on the sample and cryogenic environment, we mapped out the same area in momentum space at an elevated temperature of $T=35$~K for La-doped or at $T=40$~K for Nd-doped samples. According to earlier measurements \cite{FriemelLi12}, the quasielastic signal is suppressed at this temperature below the detection limit, so by using the high-temperature datasets as background, we could obtain clean momentum-space distributions of the magnetic intensity by subtraction. The data were analyzed and symmetrized using \textsc{Dave} \cite{AzuahKneller09} and \textsc{Horace} \cite{EwingsButs16} software.

\vspace{-3pt}\subsection{Magnetic phase diagram of \cnb}\vspace{-2pt}

The La-substituted cerium hexaborides \clb\ have been a subject of active research for a long time \cite{CameronFriemel16, SeraSato87, TomitaKunii92, TayamaSakakibara97, KobayashiSera00, NakamuraEndo06, NagaoIgarashi06, SchenckGygax07, LoveseyFernandezRodriguez07, KuwaharaIwasa07, KondoTou07, KuwaharaIwasa09, KunimoriTanida10, AnisimovGlushkov13, MatsumuraMichimura14, FriemelJang15, JangPortnichenko17}, and complete characterization of all La substituted \clb\ single crystals, which we used in this work, was already done previously \cite{FriemelJang15, JangPortnichenko17}. However, there have been only very few works devoted to the magnetic phase diagrams of Nd-substituted \cnbx\ \cite{MignotRobert09, KobayashiYoshino03, YoshinoKobayashi04}. We therefore start with a detailed investigation of the magnetic phase diagram of our sample, \cnb, using specific heat and magnetization measurements.

\begin{figure}[t]
  \includegraphics[width=1\linewidth]{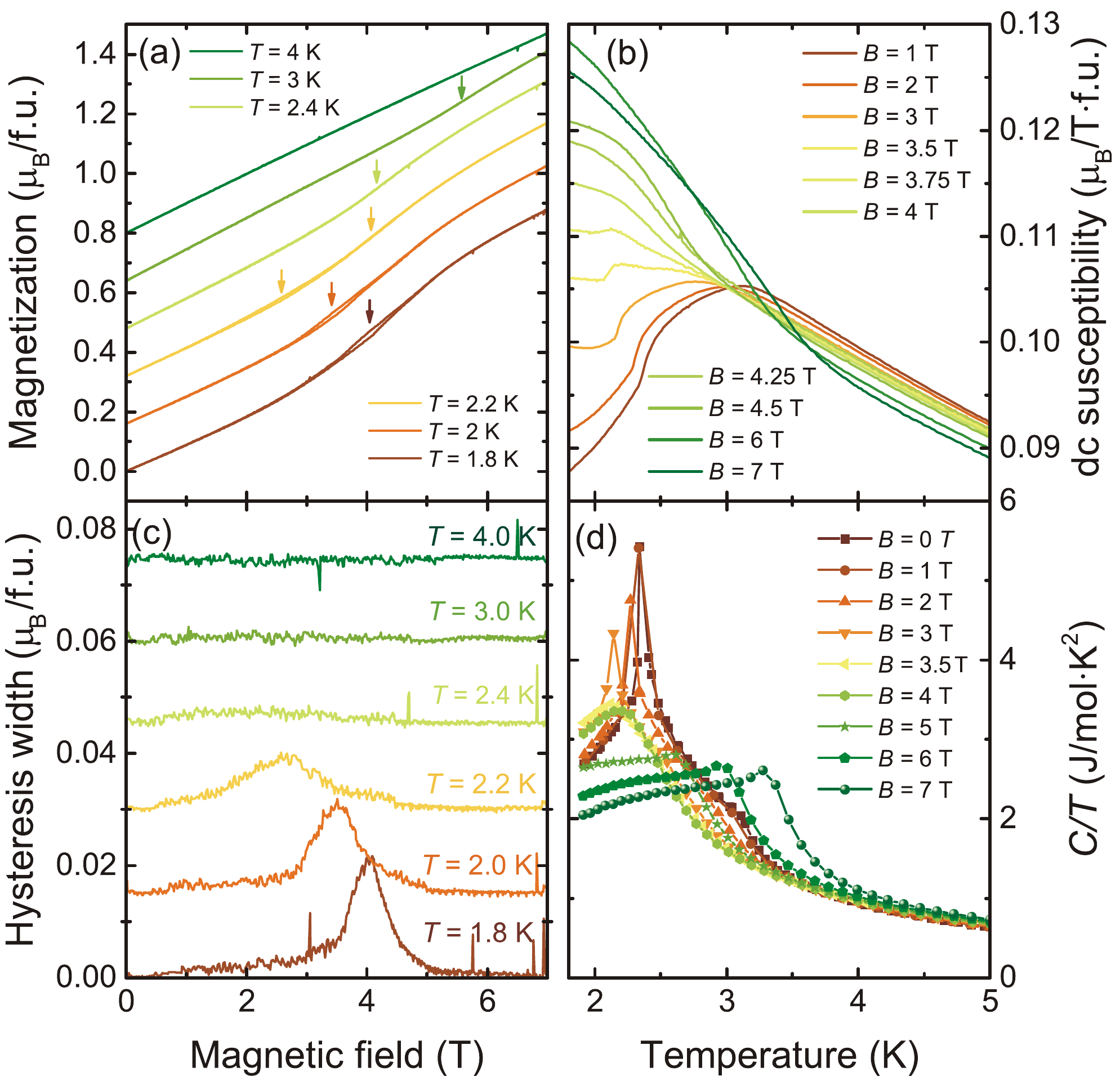}
  \caption{Magnetization and specific heat of \cnb\ measured with $\mathbf{B}\parallel[1\bar{1}0]$.
(a)~Magnetic field dependencies of magnetization. Data are offset vertically for clarity. (b)~Temperature dependencies of dc susceptibility $\chi(T) = M(T)/B$. (c)~Hysteresis width calculated as $M(B)\!\!\uparrow-~M(B)\!\!\downarrow$. Data are offset vertically for clarity. (d)~Specific heat, $C(T)/T$, measured in different magnetic fields.}
  \label{Thermodynamic}
\end{figure}

Figure~\ref{Thermodynamic} shows a summary of thermodynamic data which we used to reconstruct the magnetic-field\,--\,temperature phase diagram of \cnb. First of all, let us consider magnetic-field dependencies of magnetization measured up to $B=7$~T, which are shown in Fig.~\ref{Thermodynamic}\,(a). High-temperature ($T>3.5$~K) magnetization is exactly linear in the whole available field range, and a small change of slope appears at high fields, when sample is cooled down below $T=3.2$~K. This critical field decreases with temperature, and below $T=2.3$~K, we observe a hysteresis, which indicates a first-order phase transition from the AFQ phase~II to the AFM phase~III, consistently with previously published results on Ce$_{0.8}$Nd$_{0.2}$B$_6$ \cite{YoshinoKobayashi04}. Field dependencies of the hysteresis width, calculated by subtraction $M(B)\!\!\uparrow-~M(B)\!\!\downarrow$, are shown in Fig.~\ref{Thermodynamic}\,(c), and one can see that below $T_{\text{N2}}=2.38$~K, upon decreasing temperature, the maximum of the hysteresis shifts to higher fields and increases in amplitude.

\begin{figure}[b!]
\begin{center}\vspace{-3pt}
  \includegraphics[width=0.94\linewidth]{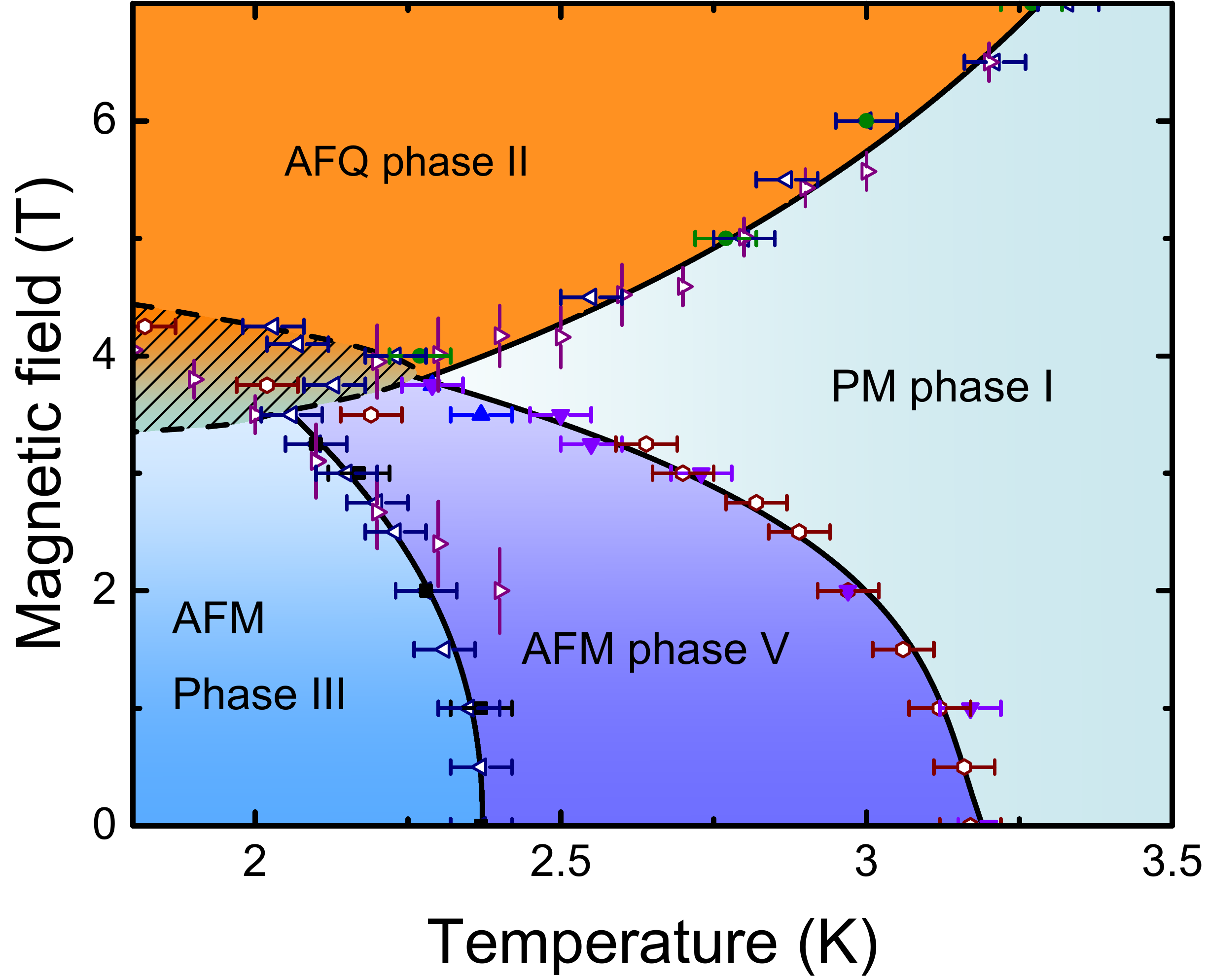}\vspace{-8pt}
\end{center}
  \caption{
  Magnetic phase diagram of \cnb, reconstructed from thermodynamic measurements. Open and filled symbols are extracted from results of magnetization and specific heat measurements, respectively. Lines are drawn as guides for the eyes. Error bars in this and all following figures represent one standard deviation.\vspace{-3pt}
  }
  \label{PhDCNB}
\end{figure}

% MvsT
Temperature dependencies of the DC magnetic susceptibility $\chi(T)=M(T)/B$ are shown in Fig.~\ref{Thermodynamic}\,(b). First of all, we discuss the low-field curves, $B<3$~T. One can clearly see a well pronounced maximum at $T_{\text{N1}}\simeq3$~K upon entering the phase~V and a kink at $T_{\text{N2}}\simeq2.38$~K, which corresponds to the III--V phase boundary. Increasing the magnetic field smears out this behaviour, and one can see almost constant $\chi(T)$ dependencies for $B=3.75$~T. The high-field phase was previously identified as the AFQ phase~II \cite{YoshinoKobayashi04}, and we extracted the boundary as a locus of minima in the $\partial\!\chi(T)/\partial{}T$ curves.

\begin{figure*}
\includegraphics[width=\linewidth]{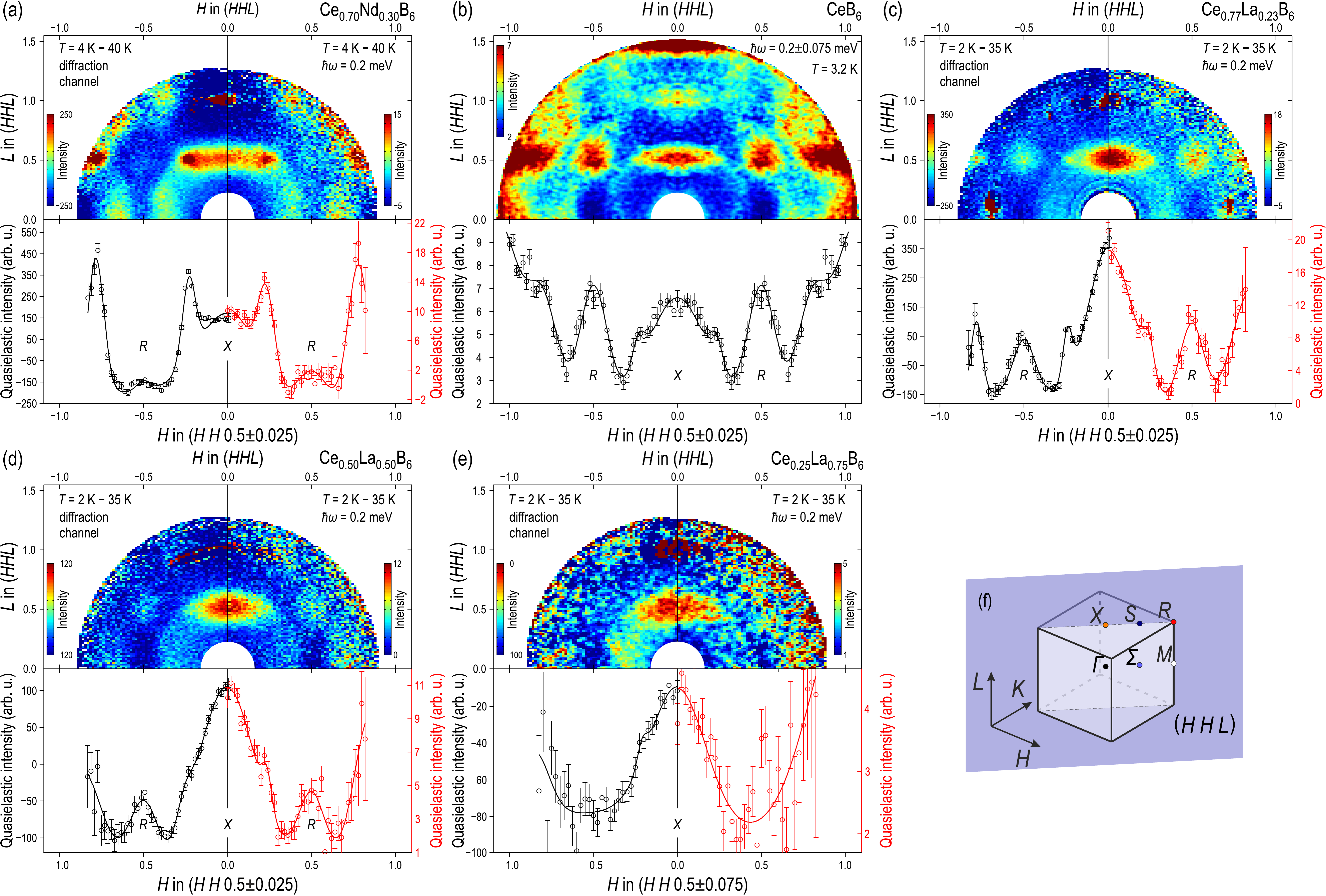}\vspace{3pt}
  \caption{Summary of INS results, measured on Ce$_{1-x}R_x$B$_6$. The top panel of every figure represents the $\mathbf{Q}$-dependence of QEMS intensity within the $(HHL)$ plane. Lower panels show cuts along the $(HH\frac{1}{2})$ direction, obtained by integration within $\pm0.025$ along the $L$ axis.
  (a)~Background-subtracted INS intensity measured on \cnb\ with the MACS spectrometer at an energy transfer $\Delta{}E=0.2$~meV and in the diffraction channel, as shown at the right and left sides of each panel, respectively. The temperatures of foreground and background measurements are shown in the corner of every panel.
  (b)~INS results for the parent compound \cb, measured with the time-of-flight spectrometer IN5 at $T=3.2$~K, integrated within an energy window $E=[0.125,0.275]$~meV.
  (c--e)~The background-subtracted data for the three samples of \clb, measured and presented in the same way as the data in panel (a).
  (f)~Schematic representation of the $(HHL)$ scattering plane in the cubic Brillouin zone of \cb\ with labeling of high-symmetry points.}
  \label{QEMS}
\end{figure*}

Similar features have been observed in the specific-heat measurements shown in Fig.~\ref{Thermodynamic}\,(d). The temperature dependence of the zero-field specific heat consists of a sharp $\lambda$-shape anomaly at $T_{\text{N}1}=2.34$~K and a small peak at $T_{\text{N}2}=3.12$~K. Application of magnetic field gradually suppresses both $T_{\text{N}1}$ and $T_{\text{N}2}$ until at a field $B=3.5$~T they merge into a single broad peak centered at $T=2.13$~K, which remains at the same position up to $B=4$~T. At higher fields, we observe a new peak, which corresponds to the I--II phase boundary.

Combining results of specific heat and magnetization measurements, we have reconstructed the magnetic-field\,--\,temperature phase diagram of \cnb, which is shown in Fig.~\ref{PhDCNB}. In spite of the perfect agreement between specific-heat and magnetization data in both high-field and low-field limits, intermediate region around $B\approx4$~T, where all 3 phase boundaries intersect, is rather ambiguous. Taking into account that the phase transition from the AFM phase~III to the field-induced AFQ phase~II is first order, we consider that this critical area corresponds to the coexistence of phases II, III and V. Our results are reasonably consistent with previously published phase diagrams of \cnbx\ \cite{KobayashiYoshino03, YoshinoKobayashi04}, and here we have presented the magnetic phase diagram for another composition with $x=0.3$, which we used in our INS measurements.\vspace{-6pt}

\vspace{-3pt}\subsection{Neutron scattering}\vspace{-2pt}

In this section we discuss our experimental approach to mapping the QEMS intensity distribution and the results of our INS measurements. The magnetic quasielastic line has its maximum of intensity near zero energy transfer, where strong nonmagnetic background intensity from incoherent scattering on the sample and sample environment is also present. However, the quasielastic line is usually broader than the energy resolution, and therefore can be measured away from the elastic line \cite{Robinson00book}. When mapping the $\mathbf{Q}$ dependence of QEMS intensity, one usually chooses a small but nonzero energy transfer to have a compromise between magnetic and incoherent intensity that maximizes the signal-to-noise ratio. This approach has been successfully applied in many earlier studies of $f\!$-electron compounds \cite{RossatMignod88, RegnaultErkelens88, SchroederAeppli98, StockertLoehneysen98, KadowakiSato04, StockSokolov11, SinghThamizhavel11, KimuraNakatsuji13}. Here, because of the specifics of the MACS spectrometer that always measures energy-integrated scattering in parallel to the spectroscopic measurements, we had a possibility to compare this method to the brute-force subtraction of the signals measured in the diffraction channel during the same acquisition time.

The main results of our neutron-scattering measurements are presented in Fig.~\ref{QEMS}. Sample compositions are shown above the corresponding panels. The top part of every panel shows QEMS intensity distributions in the $(HHL)$ scattering plane as a color map. The bottom parts show one-dimensional cuts taken through these slices along the $(HH\frac{1}{2})$ line in the reciprocal space. Figures~\ref{QEMS}\,(a,c--e) show background-subtracted data, measured at the MACS spectrometer on \cnb\ and \clb\ as described above. Every color map is split into the left and right parts to show the data obtained in the diffraction (energy-integrated) and spectroscopic ($\Delta{}E=0.2$~meV) channels. However, because pure \cb\ was measured earlier at the IN5 spectrometer without high-temperature background, Fig.~\ref{QEMS}\,(b) shows only a symmetrized QEMS signal at a finite energy transfer, integrated in the $[0.125,0.275]$~meV energy window without background subtraction.

First of all, let us discuss the QEMS map measured on the pure \cb\ [Fig.~\ref{QEMS}(b)]. For a simple antiferromagnet, just above $T_{\text{N}}$, one would expect smeared QEMS intensity concentrated around the AFM wavevectors. However, \cb\ shows a very different picture. A lot of magnetic spectral weight is concentrated at the $R$ and $\Gamma$ points, which correspond to the propagation vector of the AFQ phase and the unconventional FM mode, respectively. Also, one can see a large elliptical hump around the $X$-point that connects additional weaker peaks at the AFM wavevectors $\mathbf{q}_1=(\pm\frac{1}{4}\,\!\pm\!\frac{1}{4}\,\frac{1}{2})$, seen as a central maximum with two shoulders at the bottom of Fig.~\ref{QEMS}\,(b).

La substitution strongly suppresses the inelastic-scattering intensity at the zone center and at the AFM wavevectors $\mathbf{q}_1=(\frac{1}{4}\frac{1}{4}\frac{1}{2})$ and $\mathbf{q}_2=(\frac{1}{4}\frac{1}{4}0)$, so that it can be recognized in the QEMS maps only for the pure \cb\ and Ce$_{0.77}$La$_{0.23}$B$_6$ compounds. Remarkably, the peak at the $R$ point can be clearly seen up to a rather high La concentration of $x=0.5$ and gets completely suppressed only in the most diluted Ce$_{0.25}$La$_{0.75}$B$_6$ sample. All the spectral weight is accumulated at the $X$ point, where the intensity (per mole Ce) goes up with increasing La concentration. In highly La-diluted samples, the elliptical feature at the $X$ point dominates the QEMS intensity distribution.

Now we turn to the discussion of the Nd-substituted compound. As known from the literature \cite{MignotRobert09}, pure NdB$_6$ develops AFM order with the propagation vector $(\frac{1}{2}00)$, which coincides with the $X$ point. This order persists in the magnetic phase diagram of \cnb\ up to $x\approx0.5$ (see Fig.~\ref{PhDsub}). Knowing that both La and Nd substitutions lead to an effective hole doping and should therefore cause similar changes of the Fermi surface, it is natural to expect a strong rise of intensity at the $X$ point with simultaneous suppression of excitations at the $\Gamma$ and $R$ points also in \cnbx. At high Nd concentrations, the peak at the $X$ point would represent critical paramagnon fluctuations of phase~VI above $T_{\rm N}$, whereas on the La-rich side of the phase diagram similar fluctuations are also present, but never condense into a long-range magnetically ordered phase because of the strong dilution of the moments \cite{KoitzschHeming16}. We are therefore interested in looking at an intermediate Nd concentration, shortly before reaching the phase~VI, to see if the spectral-weight transfer to the $X$ point takes place in a similar fashion as on the La-doped side. Figure~\ref{QEMS}\,(a) shows the QEMS intensity maps measured on \cnb. Indeed, one can see a strong reduction of magnetic intensity at the $R$ point, which is consistent with the rapid suppression of the AFQ phase by Nd. Instead, we find a narrow diffuse peak, centered at the $X$ point, that connects the strong intensity maxima at the equivalent AFM wavevectors, $\mathbf{q}_1=(\pm\!\frac{1}{4}\,\pm\!\frac{1}{4}\,\frac{1}{2})$, and an elongated broad peak extending along $(\frac{1}{3}\frac{1}{3}L)$. The presence of extended peaks in momentum space with multiple local maxima of QEMS intensity is a signature of itinerant frustration in this system, which can explain the proximity of multiple AFM phases in a small region of the phase diagram.

As the last technical remark, we would like to compare the background-subtracted datasets in Figs.~\ref{QEMS}(a,c-e), measured in the diffraction (left) and spectroscopic (right) channels. It is seen that the data measured by both channels look very similar, and the signal-to-noise ratio is somewhat better for the diffraction channel. This came out as a surprise, because the magnetic signal at zero energy transfer is expected to be of approximately the same amplitude as at 0.2~meV \cite{JangFriemel14}, whereas the background level at the elastic position is higher by a factor of $\sim$\,250. However, because the diffraction channel is not restricted to elastic scattering, but integrates over all neutron energies that do not fulfill the Bragg condition on the analyzer, the amplitude of the magnetic signal is enhanced due to the broad width of the quasielastic Lorentzian peak as compared to the elastic line. Hence, our results imply that the magnetic intensity in the diffraction channel is strongly dominated by inelastic scattering, whereas the background comes predominantly from the incoherent elastic line.

\begin{figure}[t]
  \includegraphics[width=1\linewidth]{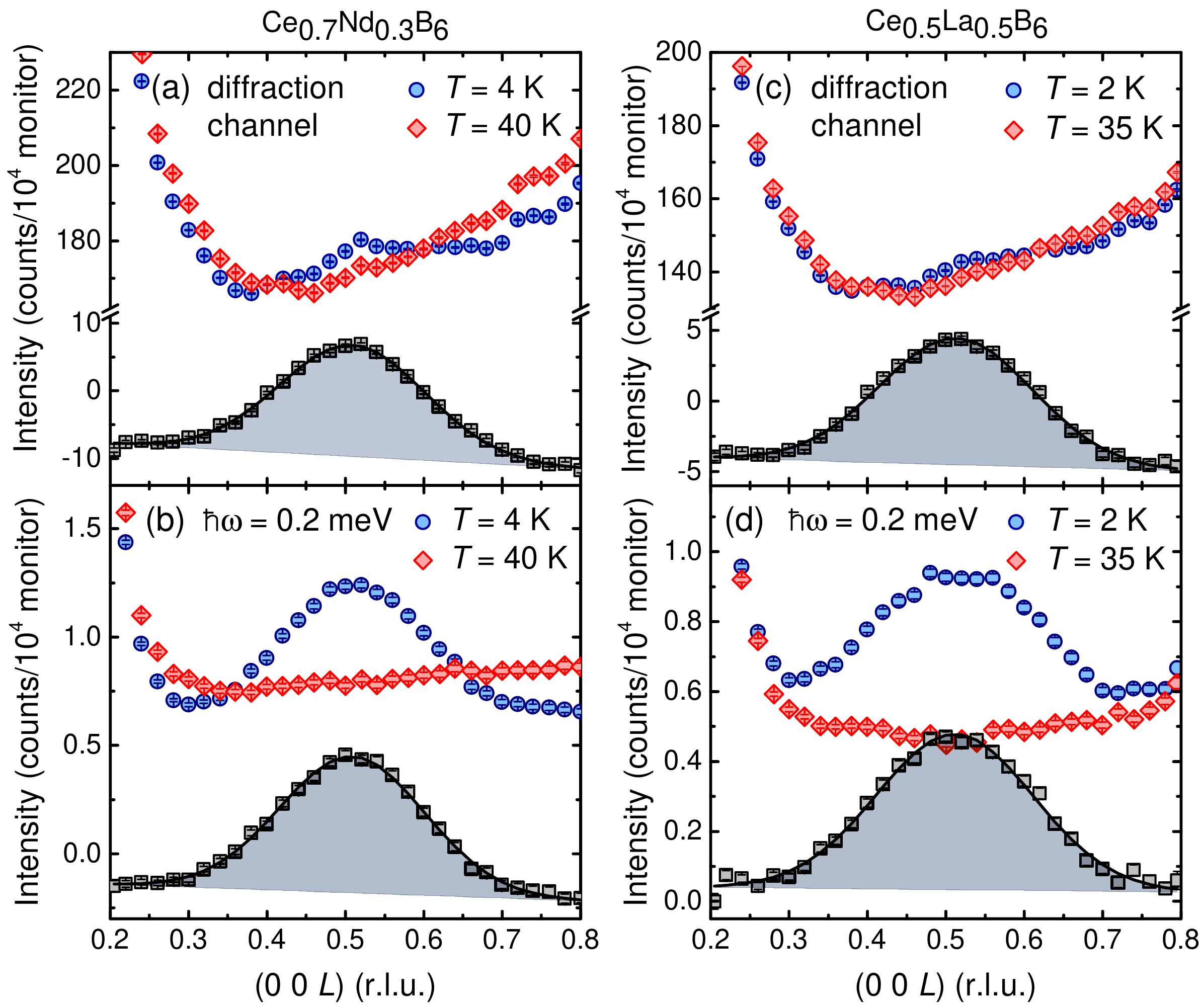}
  \caption{Intensity profiles along the $(00L)$ direction for the 30\%~Nd (left) and 50\%~La (right) doped samples. The high- and low-$T$ data are shown above the corresponding background-subtracted signal. The data in the diffraction (top) and spectroscopic (bottom) channels were obtained by integrating the MACS datasets within $-0.1 \leq H \leq 0.1$. There is on average a 50\% improvement in the signal-to-noise ratio for the diffraction channel as compared to the inelastic measurement.}
  \label{Subtraction}
\end{figure}

To put this discussion on a more quantitative ground, in Fig.~\ref{Subtraction} we show $(00L)$ intensity profiles for two of our samples, with 30\%~Nd and 50\%~La doping, obtained by integrating the data in the diffraction (top) and spectroscopic (bottom) channels within $-0.1 \leq H \leq 0.1$. We see that the diffuse magnetic peak at the $X$ point is clearly seen in the spectroscopic channel already in the raw data, with an amplitude of $\sim$\,0.5--0.6 units on top of a similarly intense background. In the diffraction channel, the peak can be barely recognized in the raw data on top of the huge background, but the subtraction shows that its amplitude is increased to $\sim$\,16 and 9 for the Nd- and La-doped samples. At the same time, the background at the peak position is increased to $\sim$\,170 and 140, respectively. However, the corresponding signal-to-noise ratios, which we estimate as the peak amplitude divided by the statistical standard deviation of the count rate near the peak maximum in the background-subtracted data, are $\sim$\,50 for the diffraction channel and $\sim$\,32 for the spectroscopic channel. In other words, in spite of the high background level, the diffraction channel provides a signal with a 1.5 times better signal-to-noise ratio. This would hold for any sample with a similarly low incoherent scattering cross-section and approximately the same quasielastic line width as those of CeB$_6$. The good agreement between the signals in both channels confirms that the spectral shape of the quasielastic line is independent of $\mathbf{Q}$ and can be described by a quasielastic Lorentzian function with a single energy width for every given temperature and doping level. We expect that setting the spectroscopic channel to zero energy transfer could improve the signal-to-noise ratio in the diffraction channel even further, as the analyzer would then act as a band-stop filter for the incoherent elastic line. We suggest this as a possible method for an effective momentum-space mapping of QEMS intensity using the MACS spectrometer, which could be especially useful for dilute samples with weak magnetism.

\vspace{-2pt}\section{Discussion and conclusions}\vspace{-1pt}

Before we start the discussion of our results, we would like to recall some experimental facts concerning the electronic structures of light rare-earth hexaborides. Previously we already argued that the propagation vectors of both AFM and AFQ phases in pure \cb\ are dictated by the nesting properties of its Fermi surface \cite{KoitzschHeming16}. In nonmagnetic LaB$_6$ and antiferromagnetic NdB$_6$, dHvA measurements found no significant difference in the size of the Fermi surface due to a strong localization of the Nd$^{3+}$~$4f$ electrons \cite{OnukiUmezawa89, ArkoCrabtree76}. The stoichiometric NdB$_6$ has an AFM structure (phase~VI) with the propagation vector $\mathbf{q}_0=(00\frac{1}{2})$, which persists in \cnbx\ about halfway across the phase diagram (see Fig.~\ref{PhDsub}). Therefore, even in the \cb\ parent compound, a remnant diffuse peak in neutron scattering is seen at the $X$ point, which we could previously associate with a Fermi-surface nesting vector by a direct comparison with ARPES data \cite{KoitzschHeming16}. Apparently, hole doping of \cb\ with either Nd or La enhances the nesting at the $X$ point. From the similarity of electronic structures in LaB$_6$ and NdB$_6$, one expects that the dominant nesting vector in nonmagnetic LaB$_6$ should also coincide with the $X$ point, even if it can no longer lead to any magnetic order due to the absence of local magnetic moments.

We want to point out that neutron scattering directly probes the imaginary part of the dynamic spin susceptibility $\chi(q,\omega)$. In linear response theory for an electron gas, it is proportional to the Lindhard function $$\chi_\mathbf{q}=\sum_{\mathbf{k}}\frac{n_\text{F}(\epsilon_{\mathbf{k}+\mathbf{q}})-n_\text{F}(\epsilon_{\mathbf{k}})}{\epsilon_{\mathbf{k}}-\epsilon_{\mathbf{k}+\mathbf{q}}},$$
where $\epsilon(\mathbf{k})$ is the dispersion relation for the conduction electrons, and $n(\epsilon)$ is the Fermi function. On the one hand, the same Lindhard function contains information about Fermi-surface nesting properties, as its real part at $\omega\rightarrow0$ is peaked at the nesting vectors and determines the propensity towards Fermi-surface instabilities in charge- or spin-density-wave systems \cite{ChanHeine73, Fawcett88, BorisenkoKordyuk08}. On the other hand, it also enters the expression for the oscillatory Ruderman-Kittel-Kasuya-Yosida (RKKY) interaction between localized Kondo spins in metals, which is mediated by the conduction electrons over long distances \cite{RudermanKittel54, Kasuya56, Yafet87, KimLee96, Aristov97, LitvinovDugaev98}. Therefore, when localized magnetic impurities are added to a nonmagnetic metal, they tend to develop short-range dynamic correlations that are seen as QEMS scattering in neutron spectroscopy or even lead to a long-range magnetic ordering of the impurity spins \cite{BrownCaudron85, FretwellDugale99}. The QEMS intensity can therefore develop maxima at the Fermi-surface nesting vectors in $\mathbf{Q}$ space even in dilute Kondo allows that are far from any ordering instability.

This general principle is nicely demonstrated by our results on the La-diluted \cb, which shows a pronounced QEMS peak at the $X$ point. As we follow the evolution of the diffuse magnetic intensity distribution from the \cb\ parent compound towards the dilute Kondo-impurity limit, we observe a continuous suppression of the critical scattering associated with the AFQ and AFM phases, peaked at the $R$, $\Gamma$, and $\mathbf{q}_{1,2}$ points, whereas the spectral weight at the $X$ point gradually accumulates.

On the other hand, substitution of Nd for Ce has a dual effect on the system. First, as already mentioned, it reduces the 4f\,--\,5d hybridization and shrinks the electron-like Fermi surfaces, that is equivalent to an effective hole doping. Second, it introduces large magnetic moments of Nd$^{3+}$ into the system, increasing its propensity towards magnetic ordering, which is an opposite effect to the nonmagnetic La$^{3+}$ dilution of the Ce$^{3+}$ moments. As follows from our results presented in Fig.~\ref{QEMS}, the evolution of the Fermi-surface nesting properties in both systems is similar, leading to an enhanced QEMS intensity near the $X$ point at the expense of the suppressed peak at the $R$ point. However, in contrast to \clb\ that tends to develop an elusive ``hidden order'' phase~IV, three distinct AFM phases are found in \cnbx, which compete in the intermediate doping range. The corresponding fluctuations are observed above $T_{\rm N}$ as extended diffuse peaks with several local maxima at the corresponding~wave~vectors.

In magnetic insulators, a variety of magnetic phases may arise from the competition among different interactions, such as frustrated Heisenberg exchange, antisymmetric Dzyaloshinskii-Moriya interaction, single-ion anisotropy, compass-type interactions, etc. This is why frustrated magnets usually exhibit complex magnetic phase diagrams in which multiple thermodynamic phases are found in close proximity \cite{RauLee14, IqbalJeschke15, BierliLhuillier16, TymoshenkoOnykiienko17, SchmidtThalmeier17}. Our present results indicate that the similarly rich phase diagram of \cnbx\ and \clb, which comprises at least two different multipolar phases and a number of AFM phases, could be the result of a similar competition among different Fermi-surface nesting vectors, where the charge-carrier doping and magnetic-moment concentration play the role of tuning parameters. It is natural to describe this situation as a typical example of itinerant frustration. Taking into account the easily accessible temperature and magnetic-field ranges, the \cnbx\ compounds provide an excellent playground for future investigations of the interplay between electronic and magnetic degrees of freedom in rare-earth hexaborides.\vspace{-2pt}

\section*{ACKNOWLEDGMENTS}\vspace{-2pt}

We thank O.~Stockert for discussions and for the critical reading of our manuscript. This work has been funded in part by the German Research Foundation (DFG) under the individual grant \mbox{IN 209/3-2}, through the Research Training Group GRK~1621, and the Collaborative Research Center SFB~1143 (project C03). The first author acknowledges support from the International Max Planck Research School for Chemistry and Physics of Quantum Materials (IMPRS-CPQM). Access to MACS was provided by the Center for High Resolution Neutron Scattering, a partnership between the National Institute of Standards and Technology and the National Science Foundation under Agreement No.~DMR-1508249.

\bibliographystyle{my-apsrev}\bibliography{CeB6_spectral}

%merlin.mbs apsrev4-1.bst 2010-07-25 4.21a (PWD, AO, DPC) hacked
%Control: key (0)
%Control: author (72) initials jnrlst
%Control: editor formatted (1) identically to author
%Control: production of article title (-1) disabled
%Control: page (0) single
%Control: year (1) truncated
%Control: production of eprint (0) enabled
\begin{thebibliography}{103}%
\makeatletter
\providecommand \@ifxundefined [1]{%
 \@ifx{#1\undefined}
}%
\providecommand \@ifnum [1]{%
 \ifnum #1\expandafter \@firstoftwo
 \else \expandafter \@secondoftwo
 \fi
}%
\providecommand \@ifx [1]{%
 \ifx #1\expandafter \@firstoftwo
 \else \expandafter \@secondoftwo
 \fi
}%
\providecommand \natexlab [1]{#1}%
\providecommand \enquote  [1]{``#1''}%
\providecommand \bibnamefont  [1]{#1}%
\providecommand \bibfnamefont [1]{#1}%
\providecommand \citenamefont [1]{#1}%
\providecommand \href@noop [0]{\@secondoftwo}%
\providecommand \href[0]{\begingroup \@sanitize@url \@href}%
\providecommand \@href[1]{\@@startlink{#1}\@@href}%
\providecommand \@@href[1]{\endgroup#1\@@endlink}%
\providecommand \@sanitize@url [0]{\catcode `\\12\catcode `\$12\catcode
  `\&12\catcode `\#12\catcode `\^12\catcode `\_12\catcode `\%12\relax}%
\providecommand \@@startlink[1]{}%
\providecommand \@@endlink[0]{}%
\providecommand \url  [0]{\begingroup\@sanitize@url \@url }%
\providecommand \@url [1]{\endgroup\@href {#1}{\urlprefix }}%
\providecommand \urlprefix  [0]{URL }%
\providecommand \Eprint [0]{\href}%
\providecommand \doibase [0]{http://dx.doi.org}%
\providecommand \selectlanguage [0]{\@gobble}%
\providecommand \bibinfo  [0]{\@secondoftwo}%
\providecommand \bibfield  [0]{\@secondoftwo}%
\providecommand \translation [1]{[#1]}%
\providecommand \BibitemOpen [0]{}%
\providecommand \bibitemStop [0]{}%
\providecommand \bibitemNoStop [0]{.\EOS\space}%
\providecommand \EOS [0]{\spacefactor3000\relax}%
\providecommand \BibitemShut  [1]{\csname bibitem#1\endcsname}%
\let\auto@bib@innerbib\@empty
%</preamble>
\bibitem [{\citenamefont {Straub}\ \emph {\textit{et~al.}}(1999)\citenamefont
  {Straub}, \citenamefont {Finteis}, \citenamefont {Claessen}, \citenamefont
  {Steiner}, \citenamefont {H\"ufner}, \citenamefont {Blaha}, \citenamefont
  {Oglesby},\ and\ \citenamefont {Bucher}}]{StraubFinteis99}%
  \BibitemOpen
  \bibfield  {author} {\bibinfo {author} {\bibfnamefont {T.}~\bibnamefont
  {Straub}}, \bibinfo {author} {\bibfnamefont {T.}~\bibnamefont {Finteis}},
  \bibinfo {author} {\bibfnamefont {R.}~\bibnamefont {Claessen}}, \bibinfo
  {author} {\bibfnamefont {P.}~\bibnamefont {Steiner}}, \bibinfo {author}
  {\bibfnamefont {S.}~\bibnamefont {H\"ufner}}, \bibinfo {author}
  {\bibfnamefont {P.}~\bibnamefont {Blaha}}, \bibinfo {author} {\bibfnamefont
  {C.~S.}\ \bibnamefont {Oglesby}}, \ and\ \bibinfo {author} {\bibfnamefont
  {E.}~\bibnamefont {Bucher}},\
  }\href{https://link.aps.org/doi/10.1103/PhysRevLett.82.4504} {\bibfield
  {journal} {\bibinfo  {journal} {Phys. Rev. Lett.}\ }\textbf {\bibinfo
  {volume} {82}},\ \bibinfo {pages} {4504} (\bibinfo {year}
  {1999})}\BibitemShut {NoStop}%
\bibitem [{\citenamefont {Sch{\"a}fer}\ \emph
  {\textit{et~al.}}(2003)\citenamefont {Sch{\"a}fer}, \citenamefont {Sing},
  \citenamefont {Claessen}, \citenamefont {Rotenberg}, \citenamefont {Zhou},
  \citenamefont {Thorne},\ and\ \citenamefont {Kevan}}]{SchaeferSing03}%
  \BibitemOpen
  \bibfield  {author} {\bibinfo {author} {\bibfnamefont {J.}~\bibnamefont
  {Sch{\"a}fer}}, \bibinfo {author} {\bibfnamefont {M.}~\bibnamefont {Sing}},
  \bibinfo {author} {\bibfnamefont {R.}~\bibnamefont {Claessen}}, \bibinfo
  {author} {\bibfnamefont {E.}~\bibnamefont {Rotenberg}}, \bibinfo {author}
  {\bibfnamefont {X.~J.}\ \bibnamefont {Zhou}}, \bibinfo {author}
  {\bibfnamefont {R.~E.}\ \bibnamefont {Thorne}}, \ and\ \bibinfo {author}
  {\bibfnamefont {S.~D.}\ \bibnamefont {Kevan}},\
  }\href{\doibase/10.1103/PhysRevLett.91.066401} {\bibfield  {journal}
  {\bibinfo  {journal} {Phys. Rev. Lett.}\ }\textbf {\bibinfo {volume} {91}},\
  \bibinfo {pages} {066401} (\bibinfo {year} {2003})}\BibitemShut {NoStop}%
\bibitem [{\citenamefont {Inosov}\ \emph {\textit{et~al.}}(2008)\citenamefont
  {Inosov}, \citenamefont {Zabolotnyy}, \citenamefont {Evtushinsky},
  \citenamefont {Kordyuk}, \citenamefont {B{\"u}chner}, \citenamefont
  {Follath}, \citenamefont {Berger},\ and\ \citenamefont
  {Borisenko}}]{InosovZabolotnyy08}%
  \BibitemOpen
  \bibfield  {author} {\bibinfo {author} {\bibfnamefont {D.~S.}\ \bibnamefont
  {Inosov}}, \bibinfo {author} {\bibfnamefont {V.~B.}\ \bibnamefont
  {Zabolotnyy}}, \bibinfo {author} {\bibfnamefont {D.~V.}\ \bibnamefont
  {Evtushinsky}}, \bibinfo {author} {\bibfnamefont {A.~A.}\ \bibnamefont
  {Kordyuk}}, \bibinfo {author} {\bibfnamefont {B.}~\bibnamefont
  {B{\"u}chner}}, \bibinfo {author} {\bibfnamefont {R.}~\bibnamefont
  {Follath}}, \bibinfo {author} {\bibfnamefont {H.}~\bibnamefont {Berger}}, \
  and\ \bibinfo {author} {\bibfnamefont {S.~V.}\ \bibnamefont {Borisenko}},\
  }\href{\doibase/10.1088/1367-2630/10/12/125027} {\bibfield  {journal}
  {\bibinfo  {journal} {New J.~Phys.}\ }\textbf {\bibinfo {volume} {10}},\
  \bibinfo {pages} {125027} (\bibinfo {year} {2008})}\BibitemShut {NoStop}%
\bibitem [{\citenamefont {Terashima}\ \emph
  {\textit{et~al.}}(2009)\citenamefont {Terashima}, \citenamefont {Sekiba},
  \citenamefont {Bowen}, \citenamefont {Nakayama}, \citenamefont {Kawahara},
  \citenamefont {Sato}, \citenamefont {Richard}, \citenamefont {Xu},
  \citenamefont {Li}, \citenamefont {Cao}, \citenamefont {Xu}, \citenamefont
  {Ding},\ and\ \citenamefont {Takahashi}}]{TerashimaSekiba09}%
  \BibitemOpen
  \bibfield  {author} {\bibinfo {author} {\bibfnamefont {K.}~\bibnamefont
  {Terashima}}, \bibinfo {author} {\bibfnamefont {Y.}~\bibnamefont {Sekiba}},
  \bibinfo {author} {\bibfnamefont {J.~H.}\ \bibnamefont {Bowen}}, \bibinfo
  {author} {\bibfnamefont {K.}~\bibnamefont {Nakayama}}, \bibinfo {author}
  {\bibfnamefont {T.}~\bibnamefont {Kawahara}}, \bibinfo {author}
  {\bibfnamefont {T.}~\bibnamefont {Sato}}, \bibinfo {author} {\bibfnamefont
  {P.}~\bibnamefont {Richard}}, \bibinfo {author} {\bibfnamefont {Y.-M.}\
  \bibnamefont {Xu}}, \bibinfo {author} {\bibfnamefont {L.~J.}\ \bibnamefont
  {Li}}, \bibinfo {author} {\bibfnamefont {G.~H.}\ \bibnamefont {Cao}},
  \bibinfo {author} {\bibfnamefont {Z.-A.}\ \bibnamefont {Xu}}, \bibinfo
  {author} {\bibfnamefont {H.}~\bibnamefont {Ding}}, \ and\ \bibinfo {author}
  {\bibfnamefont {T.}~\bibnamefont {Takahashi}},\
  }\href{\doibase/10.1073/pnas.0900469106} {\bibfield  {journal} {\bibinfo
  {journal} {Proc. Natl. Acad. Sci.}\ }\textbf {\bibinfo {volume} {106}},\
  \bibinfo {pages} {7330} (\bibinfo {year} {2009})}\BibitemShut {NoStop}%
\bibitem [{\citenamefont {Evtushinsky}\ \emph
  {\textit{et~al.}}(2010)\citenamefont {Evtushinsky}, \citenamefont {Inosov},
  \citenamefont {Urbanik}, \citenamefont {Zabolotnyy}, \citenamefont
  {Schuster}, \citenamefont {Sass}, \citenamefont {H{\"a}nke}, \citenamefont
  {Hess}, \citenamefont {B{\"u}chner}, \citenamefont {Follath}, \citenamefont
  {Reutler}, \citenamefont {Revcolevschi}, \citenamefont {Kordyuk},\ and\
  \citenamefont {Borisenko}}]{EvtushinskyInosov10}%
  \BibitemOpen
  \bibfield  {author} {\bibinfo {author} {\bibfnamefont {D.~V.}\ \bibnamefont
  {Evtushinsky}}, \bibinfo {author} {\bibfnamefont {D.~S.}\ \bibnamefont
  {Inosov}}, \bibinfo {author} {\bibfnamefont {G.}~\bibnamefont {Urbanik}},
  \bibinfo {author} {\bibfnamefont {V.~B.}\ \bibnamefont {Zabolotnyy}},
  \bibinfo {author} {\bibfnamefont {R.}~\bibnamefont {Schuster}}, \bibinfo
  {author} {\bibfnamefont {P.}~\bibnamefont {Sass}}, \bibinfo {author}
  {\bibfnamefont {T.}~\bibnamefont {H{\"a}nke}}, \bibinfo {author}
  {\bibfnamefont {C.}~\bibnamefont {Hess}}, \bibinfo {author} {\bibfnamefont
  {B.}~\bibnamefont {B{\"u}chner}}, \bibinfo {author} {\bibfnamefont
  {R.}~\bibnamefont {Follath}}, \bibinfo {author} {\bibfnamefont
  {P.}~\bibnamefont {Reutler}}, \bibinfo {author} {\bibfnamefont
  {A.}~\bibnamefont {Revcolevschi}}, \bibinfo {author} {\bibfnamefont {A.~A.}\
  \bibnamefont {Kordyuk}}, \ and\ \bibinfo {author} {\bibfnamefont {S.~V.}\
  \bibnamefont {Borisenko}},\ }\href{\doibase/10.1103/PhysRevLett.105.147201}
  {\bibfield  {journal} {\bibinfo  {journal} {Phys. Rev. Lett.}\ }\textbf
  {\bibinfo {volume} {105}},\ \bibinfo {pages} {147201} (\bibinfo {year}
  {2010})}\BibitemShut {NoStop}%
\bibitem [{\citenamefont {He}\ \emph {\textit{et~al.}}(2011)\citenamefont {He},
  \citenamefont {Fujita}, \citenamefont {Enoki}, \citenamefont {Hashimoto},
  \citenamefont {Iikubo}, \citenamefont {Mo}, \citenamefont {Yao},
  \citenamefont {Adachi}, \citenamefont {Koike}, \citenamefont {Hussain},
  \citenamefont {Shen},\ and\ \citenamefont {Yamada}}]{HeFujita11}%
  \BibitemOpen
  \bibfield  {author} {\bibinfo {author} {\bibfnamefont {R.-H.}\ \bibnamefont
  {He}}, \bibinfo {author} {\bibfnamefont {M.}~\bibnamefont {Fujita}}, \bibinfo
  {author} {\bibfnamefont {M.}~\bibnamefont {Enoki}}, \bibinfo {author}
  {\bibfnamefont {M.}~\bibnamefont {Hashimoto}}, \bibinfo {author}
  {\bibfnamefont {S.}~\bibnamefont {Iikubo}}, \bibinfo {author} {\bibfnamefont
  {S.-K.}\ \bibnamefont {Mo}}, \bibinfo {author} {\bibfnamefont
  {H.}~\bibnamefont {Yao}}, \bibinfo {author} {\bibfnamefont {T.}~\bibnamefont
  {Adachi}}, \bibinfo {author} {\bibfnamefont {Y.}~\bibnamefont {Koike}},
  \bibinfo {author} {\bibfnamefont {Z.}~\bibnamefont {Hussain}}, \bibinfo
  {author} {\bibfnamefont {Z.-X.}\ \bibnamefont {Shen}}, \ and\ \bibinfo
  {author} {\bibfnamefont {K.}~\bibnamefont {Yamada}},\
  }\href{https://link.aps.org/doi/10.1103/PhysRevLett.107.127002} {\bibfield
  {journal} {\bibinfo  {journal} {Phys. Rev. Lett.}\ }\textbf {\bibinfo
  {volume} {107}},\ \bibinfo {pages} {127002} (\bibinfo {year}
  {2011})}\BibitemShut {NoStop}%
\bibitem [{\citenamefont {Inosov}\ \emph {\textit{et~al.}}(2009)\citenamefont
  {Inosov}, \citenamefont {Evtushinsky}, \citenamefont {Koitzsch},
  \citenamefont {Zabolotnyy}, \citenamefont {Borisenko}, \citenamefont
  {Kordyuk}, \citenamefont {Frontzek}, \citenamefont {Loewenhaupt},
  \citenamefont {L{\"o}ser}, \citenamefont {Mazilu}, \citenamefont
  {Bitterlich}, \citenamefont {Behr}, \citenamefont {Hoffmann}, \citenamefont
  {Follath},\ and\ \citenamefont {B\"uchner}}]{InosovEvtushinsky09}%
  \BibitemOpen
  \bibfield  {author} {\bibinfo {author} {\bibfnamefont {D.~S.}\ \bibnamefont
  {Inosov}}, \bibinfo {author} {\bibfnamefont {D.~V.}\ \bibnamefont
  {Evtushinsky}}, \bibinfo {author} {\bibfnamefont {A.}~\bibnamefont
  {Koitzsch}}, \bibinfo {author} {\bibfnamefont {V.~B.}\ \bibnamefont
  {Zabolotnyy}}, \bibinfo {author} {\bibfnamefont {S.~V.}\ \bibnamefont
  {Borisenko}}, \bibinfo {author} {\bibfnamefont {A.~A.}\ \bibnamefont
  {Kordyuk}}, \bibinfo {author} {\bibfnamefont {M.}~\bibnamefont {Frontzek}},
  \bibinfo {author} {\bibfnamefont {M.}~\bibnamefont {Loewenhaupt}}, \bibinfo
  {author} {\bibfnamefont {W.}~\bibnamefont {L{\"o}ser}}, \bibinfo {author}
  {\bibfnamefont {I.}~\bibnamefont {Mazilu}}, \bibinfo {author} {\bibfnamefont
  {H.}~\bibnamefont {Bitterlich}}, \bibinfo {author} {\bibfnamefont
  {G.}~\bibnamefont {Behr}}, \bibinfo {author} {\bibfnamefont {J.-U.}\
  \bibnamefont {Hoffmann}}, \bibinfo {author} {\bibfnamefont {R.}~\bibnamefont
  {Follath}}, \ and\ \bibinfo {author} {\bibfnamefont {B.}~\bibnamefont
  {B\"uchner}},\ }\href{\doibase/10.1103/PhysRevLett.102.046401} {\bibfield
  {journal} {\bibinfo  {journal} {Phys. Rev. Lett.}\ }\textbf {\bibinfo
  {volume} {102}},\ \bibinfo {pages} {046401} (\bibinfo {year}
  {2009})}\BibitemShut {NoStop}%
\bibitem [{\citenamefont {Koitzsch}\ \emph {\textit{et~al.}}(2016)\citenamefont
  {Koitzsch}, \citenamefont {Heming}, \citenamefont {Knupfer}, \citenamefont
  {B{\"u}chner}, \citenamefont {Portnichenko}, \citenamefont {Dukhnenko},
  \citenamefont {Shitsevalova}, \citenamefont {Filipov}, \citenamefont {Lev},
  \citenamefont {Strocov}, \citenamefont {Ollivier},\ and\ \citenamefont
  {Inosov}}]{KoitzschHeming16}%
  \BibitemOpen
  \bibfield  {author} {\bibinfo {author} {\bibfnamefont {A.}~\bibnamefont
  {Koitzsch}}, \bibinfo {author} {\bibfnamefont {N.}~\bibnamefont {Heming}},
  \bibinfo {author} {\bibfnamefont {M.}~\bibnamefont {Knupfer}}, \bibinfo
  {author} {\bibfnamefont {B.}~\bibnamefont {B{\"u}chner}}, \bibinfo {author}
  {\bibfnamefont {P.~Y.}\ \bibnamefont {Portnichenko}}, \bibinfo {author}
  {\bibfnamefont {A.~V.}\ \bibnamefont {Dukhnenko}}, \bibinfo {author}
  {\bibfnamefont {N.~Y.}\ \bibnamefont {Shitsevalova}}, \bibinfo {author}
  {\bibfnamefont {V.~B.}\ \bibnamefont {Filipov}}, \bibinfo {author}
  {\bibfnamefont {L.~L.}\ \bibnamefont {Lev}}, \bibinfo {author} {\bibfnamefont
  {V.~N.}\ \bibnamefont {Strocov}}, \bibinfo {author} {\bibfnamefont
  {J.}~\bibnamefont {Ollivier}}, \ and\ \bibinfo {author} {\bibfnamefont
  {D.~S.}\ \bibnamefont {Inosov}},\ }\href{\doibase/10.1038/ncomms10876}
  {\bibfield  {journal} {\bibinfo  {journal} {Nature Commun.}\ }\textbf
  {\bibinfo {volume} {7}},\ \bibinfo {pages} {10876} (\bibinfo {year}
  {2016})}\BibitemShut {NoStop}%
\bibitem [{\citenamefont {Sinha}\ \emph {\textit{et~al.}}(1981)\citenamefont
  {Sinha}, \citenamefont {Lander}, \citenamefont {Shapiro},\ and\ \citenamefont
  {Vogt}}]{SinhaLander81}%
  \BibitemOpen
  \bibfield  {author} {\bibinfo {author} {\bibfnamefont {S.~K.}\ \bibnamefont
  {Sinha}}, \bibinfo {author} {\bibfnamefont {G.~H.}\ \bibnamefont {Lander}},
  \bibinfo {author} {\bibfnamefont {S.~M.}\ \bibnamefont {Shapiro}}, \ and\
  \bibinfo {author} {\bibfnamefont {O.}~\bibnamefont {Vogt}},\
  }\href{https://link.aps.org/doi/10.1103/PhysRevB.23.4556} {\bibfield
  {journal} {\bibinfo  {journal} {Phys. Rev.~B}\ }\textbf {\bibinfo {volume}
  {23}},\ \bibinfo {pages} {4556} (\bibinfo {year} {1981})}\BibitemShut
  {NoStop}%
\bibitem [{\citenamefont {Aeppli}\ \emph {\textit{et~al.}}(1986)\citenamefont
  {Aeppli}, \citenamefont {Yoshizawa}, \citenamefont {Endoh}, \citenamefont
  {Bucher}, \citenamefont {Hufnagl}, \citenamefont {Onuki},\ and\ \citenamefont
  {Komatsubara}}]{AeppliYoshizawa86}%
  \BibitemOpen
  \bibfield  {author} {\bibinfo {author} {\bibfnamefont {G.}~\bibnamefont
  {Aeppli}}, \bibinfo {author} {\bibfnamefont {H.}~\bibnamefont {Yoshizawa}},
  \bibinfo {author} {\bibfnamefont {Y.}~\bibnamefont {Endoh}}, \bibinfo
  {author} {\bibfnamefont {E.}~\bibnamefont {Bucher}}, \bibinfo {author}
  {\bibfnamefont {J.}~\bibnamefont {Hufnagl}}, \bibinfo {author} {\bibfnamefont
  {Y.}~\bibnamefont {Onuki}}, \ and\ \bibinfo {author} {\bibfnamefont
  {T.}~\bibnamefont {Komatsubara}},\
  }\href{\doibase/10.1103/PhysRevLett.57.122} {\bibfield  {journal} {\bibinfo
  {journal} {Phys. Rev. Lett.}\ }\textbf {\bibinfo {volume} {57}},\ \bibinfo
  {pages} {122} (\bibinfo {year} {1986})}\BibitemShut {NoStop}%
\bibitem [{\citenamefont {Stockert}\ \emph {\textit{et~al.}}(2004)\citenamefont
  {Stockert}, \citenamefont {Faulhaber}, \citenamefont {Zwicknagl},
  \citenamefont {St\"u\ss{}er}, \citenamefont {Jeevan}, \citenamefont {Deppe},
  \citenamefont {Borth}, \citenamefont {K\"uchler}, \citenamefont
  {Loewenhaupt}, \citenamefont {Geibel},\ and\ \citenamefont
  {Steglich}}]{StockertFaulhaber04}%
  \BibitemOpen
  \bibfield  {author} {\bibinfo {author} {\bibfnamefont {O.}~\bibnamefont
  {Stockert}}, \bibinfo {author} {\bibfnamefont {E.}~\bibnamefont {Faulhaber}},
  \bibinfo {author} {\bibfnamefont {G.}~\bibnamefont {Zwicknagl}}, \bibinfo
  {author} {\bibfnamefont {N.}~\bibnamefont {St\"u\ss{}er}}, \bibinfo {author}
  {\bibfnamefont {H.~S.}\ \bibnamefont {Jeevan}}, \bibinfo {author}
  {\bibfnamefont {M.}~\bibnamefont {Deppe}}, \bibinfo {author} {\bibfnamefont
  {R.}~\bibnamefont {Borth}}, \bibinfo {author} {\bibfnamefont
  {R.}~\bibnamefont {K\"uchler}}, \bibinfo {author} {\bibfnamefont
  {M.}~\bibnamefont {Loewenhaupt}}, \bibinfo {author} {\bibfnamefont
  {C.}~\bibnamefont {Geibel}}, \ and\ \bibinfo {author} {\bibfnamefont
  {F.}~\bibnamefont {Steglich}},\
  }\href{https://link.aps.org/doi/10.1103/PhysRevLett.92.136401} {\bibfield
  {journal} {\bibinfo  {journal} {Phys. Rev. Lett.}\ }\textbf {\bibinfo
  {volume} {92}},\ \bibinfo {pages} {136401} (\bibinfo {year}
  {2004})}\BibitemShut {NoStop}%
\bibitem [{\citenamefont {Kadowaki}\ \emph {\textit{et~al.}}(2004)\citenamefont
  {Kadowaki}, \citenamefont {Sato},\ and\ \citenamefont
  {Kawarazaki}}]{KadowakiSato04}%
  \BibitemOpen
  \bibfield  {author} {\bibinfo {author} {\bibfnamefont {H.}~\bibnamefont
  {Kadowaki}}, \bibinfo {author} {\bibfnamefont {M.}~\bibnamefont {Sato}}, \
  and\ \bibinfo {author} {\bibfnamefont {S.}~\bibnamefont {Kawarazaki}},\
  }\href{https://link.aps.org/doi/10.1103/PhysRevLett.92.097204} {\bibfield
  {journal} {\bibinfo  {journal} {Phys. Rev. Lett.}\ }\textbf {\bibinfo
  {volume} {92}},\ \bibinfo {pages} {097204} (\bibinfo {year}
  {2004})}\BibitemShut {NoStop}%
\bibitem [{\citenamefont {Wiebe}\ \emph {\textit{et~al.}}(2007)\citenamefont
  {Wiebe}, \citenamefont {Janik}, \citenamefont {MacDougall}, \citenamefont
  {Luke}, \citenamefont {Garrett}, \citenamefont {Zhou}, \citenamefont {Jo},
  \citenamefont {Balicas}, \citenamefont {Qiu}, \citenamefont {Copley},
  \citenamefont {Yamani},\ and\ \citenamefont {Buyers}}]{WiebeJanik07}%
  \BibitemOpen
  \bibfield  {author} {\bibinfo {author} {\bibfnamefont {C.~R.}\ \bibnamefont
  {Wiebe}}, \bibinfo {author} {\bibfnamefont {J.~A.}\ \bibnamefont {Janik}},
  \bibinfo {author} {\bibfnamefont {G.~J.}\ \bibnamefont {MacDougall}},
  \bibinfo {author} {\bibfnamefont {G.~M.}\ \bibnamefont {Luke}}, \bibinfo
  {author} {\bibfnamefont {J.~D.}\ \bibnamefont {Garrett}}, \bibinfo {author}
  {\bibfnamefont {H.~D.}\ \bibnamefont {Zhou}}, \bibinfo {author}
  {\bibfnamefont {Y.-J.}\ \bibnamefont {Jo}}, \bibinfo {author} {\bibfnamefont
  {L.}~\bibnamefont {Balicas}}, \bibinfo {author} {\bibfnamefont
  {Y.}~\bibnamefont {Qiu}}, \bibinfo {author} {\bibfnamefont {J.~R.~D.}\
  \bibnamefont {Copley}}, \bibinfo {author} {\bibfnamefont {Z.}~\bibnamefont
  {Yamani}}, \ and\ \bibinfo {author} {\bibfnamefont {W.~J.~L.}\ \bibnamefont
  {Buyers}},\ }\href{\doibase/10.1038/nphys522} {\bibfield  {journal} {\bibinfo
   {journal} {Nature Phys.}\ }\textbf {\bibinfo {volume} {3}},\ \bibinfo
  {pages} {96} (\bibinfo {year} {2007})}\BibitemShut {NoStop}%
\bibitem [{\citenamefont {Stock}\ \emph {\textit{et~al.}}(2012)\citenamefont
  {Stock}, \citenamefont {Broholm}, \citenamefont {Demmel}, \citenamefont
  {Van~Duijn}, \citenamefont {Taylor}, \citenamefont {Kang}, \citenamefont
  {Hu},\ and\ \citenamefont {Petrovic}}]{StockBroholm12}%
  \BibitemOpen
  \bibfield  {author} {\bibinfo {author} {\bibfnamefont {C.}~\bibnamefont
  {Stock}}, \bibinfo {author} {\bibfnamefont {C.}~\bibnamefont {Broholm}},
  \bibinfo {author} {\bibfnamefont {F.}~\bibnamefont {Demmel}}, \bibinfo
  {author} {\bibfnamefont {J.}~\bibnamefont {Van~Duijn}}, \bibinfo {author}
  {\bibfnamefont {J.~W.}\ \bibnamefont {Taylor}}, \bibinfo {author}
  {\bibfnamefont {H.~J.}\ \bibnamefont {Kang}}, \bibinfo {author}
  {\bibfnamefont {R.}~\bibnamefont {Hu}}, \ and\ \bibinfo {author}
  {\bibfnamefont {C.}~\bibnamefont {Petrovic}},\
  }\href{https://link.aps.org/doi/10.1103/PhysRevLett.109.127201} {\bibfield
  {journal} {\bibinfo  {journal} {Phys. Rev. Lett.}\ }\textbf {\bibinfo
  {volume} {109}},\ \bibinfo {pages} {127201} (\bibinfo {year}
  {2012})}\BibitemShut {NoStop}%
\bibitem [{\citenamefont {Portnichenko}\ \emph
  {\textit{et~al.}}(2015)\citenamefont {Portnichenko}, \citenamefont {Cameron},
  \citenamefont {Surmach}, \citenamefont {Deen}, \citenamefont {Paschen},
  \citenamefont {Prokofiev}, \citenamefont {Mignot}, \citenamefont {Strydom},
  \citenamefont {Telling}, \citenamefont {Podlesnyak},\ and\ \citenamefont
  {Inosov}}]{PortnichenkoCameron15}%
  \BibitemOpen
  \bibfield  {author} {\bibinfo {author} {\bibfnamefont {P.~Y.}\ \bibnamefont
  {Portnichenko}}, \bibinfo {author} {\bibfnamefont {A.~S.}\ \bibnamefont
  {Cameron}}, \bibinfo {author} {\bibfnamefont {M.~A.}\ \bibnamefont
  {Surmach}}, \bibinfo {author} {\bibfnamefont {P.~P.}\ \bibnamefont {Deen}},
  \bibinfo {author} {\bibfnamefont {S.}~\bibnamefont {Paschen}}, \bibinfo
  {author} {\bibfnamefont {A.}~\bibnamefont {Prokofiev}}, \bibinfo {author}
  {\bibfnamefont {J.-M.}\ \bibnamefont {Mignot}}, \bibinfo {author}
  {\bibfnamefont {A.~M.}\ \bibnamefont {Strydom}}, \bibinfo {author}
  {\bibfnamefont {M.~T.~F.}\ \bibnamefont {Telling}}, \bibinfo {author}
  {\bibfnamefont {A.}~\bibnamefont {Podlesnyak}}, \ and\ \bibinfo {author}
  {\bibfnamefont {D.~S.}\ \bibnamefont {Inosov}},\
  }\href{https://link.aps.org/doi/10.1103/PhysRevB.91.094412} {\bibfield
  {journal} {\bibinfo  {journal} {Phys. Rev.~B}\ }\textbf {\bibinfo {volume}
  {91}},\ \bibinfo {pages} {094412} (\bibinfo {year} {2015})}\BibitemShut
  {NoStop}%
\bibitem [{\citenamefont {Butch}\ \emph {\textit{et~al.}}(2015)\citenamefont
  {Butch}, \citenamefont {Manley}, \citenamefont {Jeffries}, \citenamefont
  {Janoschek}, \citenamefont {Huang}, \citenamefont {Maple}, \citenamefont
  {Said}, \citenamefont {Leu},\ and\ \citenamefont {Lynn}}]{ButchManley15}%
  \BibitemOpen
  \bibfield  {author} {\bibinfo {author} {\bibfnamefont {N.~P.}\ \bibnamefont
  {Butch}}, \bibinfo {author} {\bibfnamefont {M.~E.}\ \bibnamefont {Manley}},
  \bibinfo {author} {\bibfnamefont {J.~R.}\ \bibnamefont {Jeffries}}, \bibinfo
  {author} {\bibfnamefont {M.}~\bibnamefont {Janoschek}}, \bibinfo {author}
  {\bibfnamefont {K.}~\bibnamefont {Huang}}, \bibinfo {author} {\bibfnamefont
  {M.~B.}\ \bibnamefont {Maple}}, \bibinfo {author} {\bibfnamefont {A.~H.}\
  \bibnamefont {Said}}, \bibinfo {author} {\bibfnamefont {B.~M.}\ \bibnamefont
  {Leu}}, \ and\ \bibinfo {author} {\bibfnamefont {J.~W.}\ \bibnamefont
  {Lynn}},\ }\href{\doibase/10.1103/PhysRevB.91.035128} {\bibfield  {journal}
  {\bibinfo  {journal} {Phys. Rev.~B}\ }\textbf {\bibinfo {volume} {91}},\
  \bibinfo {pages} {035128} (\bibinfo {year} {2015})}\BibitemShut {NoStop}%
\bibitem [{\citenamefont {Steglich}\ \emph {\textit{et~al.}}(1984)\citenamefont
  {Steglich}, \citenamefont {Bredl}, \citenamefont {Lieke}, \citenamefont
  {Rauchschwalbe},\ and\ \citenamefont {Sparn}}]{SteglichBredl84}%
  \BibitemOpen
  \bibfield  {author} {\bibinfo {author} {\bibfnamefont {F.}~\bibnamefont
  {Steglich}}, \bibinfo {author} {\bibfnamefont {C.~D.}\ \bibnamefont {Bredl}},
  \bibinfo {author} {\bibfnamefont {W.}~\bibnamefont {Lieke}}, \bibinfo
  {author} {\bibfnamefont {U.}~\bibnamefont {Rauchschwalbe}}, \ and\ \bibinfo
  {author} {\bibfnamefont {G.}~\bibnamefont {Sparn}},\
  }\href{\doibase/10.1016/0378-4363(84)90148-7} {\bibfield  {journal} {\bibinfo
   {journal} {Physica B+C}\ }\textbf {\bibinfo {volume} {126}},\ \bibinfo
  {pages} {82} (\bibinfo {year} {1984})}\BibitemShut {NoStop}%
\bibitem [{\citenamefont {Schlabitz}\ \emph
  {\textit{et~al.}}(1986)\citenamefont {Schlabitz}, \citenamefont {Baumann},
  \citenamefont {Pollit}, \citenamefont {Rauchschwalbe}, \citenamefont {Mayer},
  \citenamefont {Ahlheim},\ and\ \citenamefont {Bredl}}]{SchlabitzBaumann86}%
  \BibitemOpen
  \bibfield  {author} {\bibinfo {author} {\bibfnamefont {W.}~\bibnamefont
  {Schlabitz}}, \bibinfo {author} {\bibfnamefont {J.}~\bibnamefont {Baumann}},
  \bibinfo {author} {\bibfnamefont {B.}~\bibnamefont {Pollit}}, \bibinfo
  {author} {\bibfnamefont {U.}~\bibnamefont {Rauchschwalbe}}, \bibinfo {author}
  {\bibfnamefont {H.~M.}\ \bibnamefont {Mayer}}, \bibinfo {author}
  {\bibfnamefont {U.}~\bibnamefont {Ahlheim}}, \ and\ \bibinfo {author}
  {\bibfnamefont {C.~D.}\ \bibnamefont {Bredl}},\ }in\
  \href{\doibase/10.1007/978-94-011-1622-0_9} {\emph {\bibinfo {booktitle} {Ten
  Years of Superconductivity: 1980--1990}}}\ (\bibinfo  {publisher}
  {Springer},\ \bibinfo {year} {1986})\ pp.\ \bibinfo {pages}
  {89--95}\BibitemShut {NoStop}%
\bibitem [{\citenamefont {Zhang}\ and\ \citenamefont
  {Shelton}(2000)}]{ZhangShelton00book}%
  \BibitemOpen
  \bibfield  {author} {\bibinfo {author} {\bibfnamefont {L.}~\bibnamefont
  {Zhang}}\ and\ \bibinfo {author} {\bibfnamefont {R.~N.}\ \bibnamefont
  {Shelton}},\ }in\ \href@noop {} {\emph {\bibinfo {booktitle} {Magnetism in
  Heavy Fermion Systems}}},\ \bibinfo {editor} {edited by\ \bibinfo {editor}
  {\bibfnamefont {H.~B.}\ \bibnamefont {Radousky}}}\ (\bibinfo  {publisher}
  {World Scientific, Singapore},\ \bibinfo {year} {2000})\ Chap.~\bibinfo
  {chapter} {2}, pp.\ \bibinfo {pages} {11--146}\BibitemShut {NoStop}%
\bibitem [{\citenamefont {Pfleiderer}(2009)}]{Pfleiderer09}%
  \BibitemOpen
  \bibfield  {author} {\bibinfo {author} {\bibfnamefont {C.}~\bibnamefont
  {Pfleiderer}},\ }\href{https://link.aps.org/doi/10.1103/RevModPhys.81.1551}
  {\bibfield  {journal} {\bibinfo  {journal} {Rev. Mod. Phys.}\ }\textbf
  {\bibinfo {volume} {81}},\ \bibinfo {pages} {1551} (\bibinfo {year}
  {2009})}\BibitemShut {NoStop}%
\bibitem [{\citenamefont {Stewart}(2017)}]{Stewart17}%
  \BibitemOpen
  \bibfield  {author} {\bibinfo {author} {\bibfnamefont {G.~R.}\ \bibnamefont
  {Stewart}},\ }\href{\doibase/10.1080/00018732.2017.1331615} {\bibfield
  {journal} {\bibinfo  {journal} {Adv. Phys.}\ }\textbf {\bibinfo {volume}
  {66}},\ \bibinfo {pages} {75} (\bibinfo {year} {2017})}\BibitemShut {NoStop}%
\bibitem [{\citenamefont {Si}\ and\ \citenamefont
  {Steglich}(2010)}]{SiSteglich10}%
  \BibitemOpen
  \bibfield  {author} {\bibinfo {author} {\bibfnamefont {Q.}~\bibnamefont
  {Si}}\ and\ \bibinfo {author} {\bibfnamefont {F.}~\bibnamefont {Steglich}},\
  }\href{\doibase/10.1126/science.1191195} {\bibfield  {journal} {\bibinfo
  {journal} {Science}\ }\textbf {\bibinfo {volume} {329}},\ \bibinfo {pages}
  {1161} (\bibinfo {year} {2010})}\BibitemShut {NoStop}%
\bibitem [{\citenamefont {Stockert}\ and\ \citenamefont
  {Steglich}(2011)}]{StockertSteglich11}%
  \BibitemOpen
  \bibfield  {author} {\bibinfo {author} {\bibfnamefont {O.}~\bibnamefont
  {Stockert}}\ and\ \bibinfo {author} {\bibfnamefont {F.}~\bibnamefont
  {Steglich}},\ }\href{\doibase/10.1146/annurev-conmatphys-062910-140546}
  {\bibfield  {journal} {\bibinfo  {journal} {Annu. Rev. Condens. Matter
  Phys.}\ }\textbf {\bibinfo {volume} {2}},\ \bibinfo {pages} {79} (\bibinfo
  {year} {2011})}\BibitemShut {NoStop}%
\bibitem [{\citenamefont {Sakai}\ \emph {\textit{et~al.}}(2005)\citenamefont
  {Sakai}, \citenamefont {Shiina},\ and\ \citenamefont
  {Shiba}}]{SakaiShiina05}%
  \BibitemOpen
  \bibfield  {author} {\bibinfo {author} {\bibfnamefont {O.}~\bibnamefont
  {Sakai}}, \bibinfo {author} {\bibfnamefont {R.}~\bibnamefont {Shiina}}, \
  and\ \bibinfo {author} {\bibfnamefont {H.}~\bibnamefont {Shiba}},\
  }\href{https://doi.org/10.1143/jpsj.74.457} {\bibfield  {journal} {\bibinfo
  {journal} {J.~Phys. Soc. Jpn.}\ }\textbf {\bibinfo {volume} {74}},\ \bibinfo
  {pages} {457} (\bibinfo {year} {2005})}\BibitemShut {NoStop}%
\bibitem [{\citenamefont {Nagao}\ and\ \citenamefont
  {Igarashi}(2006)}]{NagaoIgarashi06}%
  \BibitemOpen
  \bibfield  {author} {\bibinfo {author} {\bibfnamefont {T.}~\bibnamefont
  {Nagao}}\ and\ \bibinfo {author} {\bibfnamefont {J.-i.}\ \bibnamefont
  {Igarashi}},\ }\href{\doibase/10.1103/PhysRevB.74.104404} {\bibfield
  {journal} {\bibinfo  {journal} {Phys. Rev.~B}\ }\textbf {\bibinfo {volume}
  {74}},\ \bibinfo {pages} {104404} (\bibinfo {year} {2006})}\BibitemShut
  {NoStop}%
\bibitem [{\citenamefont {Santini}\ \emph {\textit{et~al.}}(2009)\citenamefont
  {Santini}, \citenamefont {Carretta}, \citenamefont {Amoretti}, \citenamefont
  {Caciuffo}, \citenamefont {Magnani},\ and\ \citenamefont
  {Lander}}]{SantiniCarretta09}%
  \BibitemOpen
  \bibfield  {author} {\bibinfo {author} {\bibfnamefont {P.}~\bibnamefont
  {Santini}}, \bibinfo {author} {\bibfnamefont {S.}~\bibnamefont {Carretta}},
  \bibinfo {author} {\bibfnamefont {G.}~\bibnamefont {Amoretti}}, \bibinfo
  {author} {\bibfnamefont {R.}~\bibnamefont {Caciuffo}}, \bibinfo {author}
  {\bibfnamefont {N.}~\bibnamefont {Magnani}}, \ and\ \bibinfo {author}
  {\bibfnamefont {G.~H.}\ \bibnamefont {Lander}},\
  }\href{\doibase/10.1103/RevModPhys.81.807} {\bibfield  {journal} {\bibinfo
  {journal} {Rev. Mod. Phys.}\ }\textbf {\bibinfo {volume} {81}},\ \bibinfo
  {pages} {807} (\bibinfo {year} {2009})}\BibitemShut {NoStop}%
\bibitem [{\citenamefont {Paschen}\ and\ \citenamefont
  {Larrea}(2014)}]{PaschenLarrea14}%
  \BibitemOpen
  \bibfield  {author} {\bibinfo {author} {\bibfnamefont {S.}~\bibnamefont
  {Paschen}}\ and\ \bibinfo {author} {\bibfnamefont {J.}~\bibnamefont
  {Larrea}},\ }\href{https://doi.org/10.7566/jpsj.83.061004} {\bibfield
  {journal} {\bibinfo  {journal} {J.~Phys. Soc. Jpn.}\ }\textbf {\bibinfo
  {volume} {83}},\ \bibinfo {pages} {061004} (\bibinfo {year}
  {2014})}\BibitemShut {NoStop}%
\bibitem [{\citenamefont {Portnichenko}\ \emph
  {\textit{et~al.}}(2016)\citenamefont {Portnichenko}, \citenamefont {Paschen},
  \citenamefont {Prokofiev}, \citenamefont {Vojta}, \citenamefont {Cameron},
  \citenamefont {Mignot}, \citenamefont {Ivanov},\ and\ \citenamefont
  {Inosov}}]{PortnichenkoPaschen16}%
  \BibitemOpen
  \bibfield  {author} {\bibinfo {author} {\bibfnamefont {P.~Y.}\ \bibnamefont
  {Portnichenko}}, \bibinfo {author} {\bibfnamefont {S.}~\bibnamefont
  {Paschen}}, \bibinfo {author} {\bibfnamefont {A.}~\bibnamefont {Prokofiev}},
  \bibinfo {author} {\bibfnamefont {M.}~\bibnamefont {Vojta}}, \bibinfo
  {author} {\bibfnamefont {A.~S.}\ \bibnamefont {Cameron}}, \bibinfo {author}
  {\bibfnamefont {J.-M.}\ \bibnamefont {Mignot}}, \bibinfo {author}
  {\bibfnamefont {A.}~\bibnamefont {Ivanov}}, \ and\ \bibinfo {author}
  {\bibfnamefont {D.~S.}\ \bibnamefont {Inosov}},\
  }\href{https://link.aps.org/doi/10.1103/PhysRevB.94.245132} {\bibfield
  {journal} {\bibinfo  {journal} {Phys. Rev. B}\ }\textbf {\bibinfo {volume}
  {94}},\ \bibinfo {pages} {245132} (\bibinfo {year} {2016})}\BibitemShut
  {NoStop}%
\bibitem [{\citenamefont {Effantin}\ \emph {\textit{et~al.}}(1985)\citenamefont
  {Effantin}, \citenamefont {Rossat-Mignod}, \citenamefont {Burlet},
  \citenamefont {Bartholin}, \citenamefont {Kunii},\ and\ \citenamefont
  {Kasuya}}]{EffantinRossatMignod85}%
  \BibitemOpen
  \bibfield  {author} {\bibinfo {author} {\bibfnamefont {J.~M.}\ \bibnamefont
  {Effantin}}, \bibinfo {author} {\bibfnamefont {J.}~\bibnamefont
  {Rossat-Mignod}}, \bibinfo {author} {\bibfnamefont {P.}~\bibnamefont
  {Burlet}}, \bibinfo {author} {\bibfnamefont {H.}~\bibnamefont {Bartholin}},
  \bibinfo {author} {\bibfnamefont {S.}~\bibnamefont {Kunii}}, \ and\ \bibinfo
  {author} {\bibfnamefont {T.}~\bibnamefont {Kasuya}},\
  }\href{\doibase/10.1016/0304-8853(85)90382-8} {\bibfield  {journal} {\bibinfo
   {journal} {J. Magn. Magn. Mater.}\ }\textbf {\bibinfo {volume} {47}},\
  \bibinfo {pages} {145} (\bibinfo {year} {1985})}\BibitemShut {NoStop}%
\bibitem [{\citenamefont {Sluchanko}\ \emph
  {\textit{et~al.}}(2007)\citenamefont {Sluchanko}, \citenamefont {Bogach},
  \citenamefont {Glushkov}, \citenamefont {Demishev}, \citenamefont {Ivanov},
  \citenamefont {Ignatov}, \citenamefont {Kuznetsov}, \citenamefont {Samarin},
  \citenamefont {Semeno},\ and\ \citenamefont
  {Shitsevalova}}]{SluchankoBogach07}%
  \BibitemOpen
  \bibfield  {author} {\bibinfo {author} {\bibfnamefont {N.~E.}\ \bibnamefont
  {Sluchanko}}, \bibinfo {author} {\bibfnamefont {A.~V.}\ \bibnamefont
  {Bogach}}, \bibinfo {author} {\bibfnamefont {V.~V.}\ \bibnamefont
  {Glushkov}}, \bibinfo {author} {\bibfnamefont {S.~V.}\ \bibnamefont
  {Demishev}}, \bibinfo {author} {\bibfnamefont {V.~Y.}\ \bibnamefont
  {Ivanov}}, \bibinfo {author} {\bibfnamefont {M.~I.}\ \bibnamefont {Ignatov}},
  \bibinfo {author} {\bibfnamefont {A.~V.}\ \bibnamefont {Kuznetsov}}, \bibinfo
  {author} {\bibfnamefont {N.~A.}\ \bibnamefont {Samarin}}, \bibinfo {author}
  {\bibfnamefont {A.~V.}\ \bibnamefont {Semeno}}, \ and\ \bibinfo {author}
  {\bibfnamefont {N.~Y.}\ \bibnamefont {Shitsevalova}},\
  }\href{\doibase/10.1134/S106377610701013X} {\bibfield  {journal} {\bibinfo
  {journal} {J.~Exp. Theor. Phys.}\ }\textbf {\bibinfo {volume} {104}},\
  \bibinfo {pages} {120} (\bibinfo {year} {2007})}\BibitemShut {NoStop}%
\bibitem [{\citenamefont {Cameron}\ \emph {\textit{et~al.}}(2016)\citenamefont
  {Cameron}, \citenamefont {Friemel},\ and\ \citenamefont
  {Inosov}}]{CameronFriemel16}%
  \BibitemOpen
  \bibfield  {author} {\bibinfo {author} {\bibfnamefont {A.~S.}\ \bibnamefont
  {Cameron}}, \bibinfo {author} {\bibfnamefont {G.}~\bibnamefont {Friemel}}, \
  and\ \bibinfo {author} {\bibfnamefont {D.~S.}\ \bibnamefont {Inosov}},\
  }\href{\doibase/10.1088/0034-4885/79/6/066502} {\bibfield  {journal}
  {\bibinfo  {journal} {Rep. Prog. Phys.}\ }\textbf {\bibinfo {volume} {79}},\
  \bibinfo {pages} {066502} (\bibinfo {year} {2016})}\BibitemShut {NoStop}%
\bibitem [{\citenamefont {Zaharko}\ \emph {\textit{et~al.}}(2003)\citenamefont
  {Zaharko}, \citenamefont {Fischer}, \citenamefont {Schenck}, \citenamefont
  {Kunii}, \citenamefont {Brown}, \citenamefont {Tasset},\ and\ \citenamefont
  {Hansen}}]{ZaharkoFischer03}%
  \BibitemOpen
  \bibfield  {author} {\bibinfo {author} {\bibfnamefont {O.}~\bibnamefont
  {Zaharko}}, \bibinfo {author} {\bibfnamefont {P.}~\bibnamefont {Fischer}},
  \bibinfo {author} {\bibfnamefont {A.}~\bibnamefont {Schenck}}, \bibinfo
  {author} {\bibfnamefont {S.}~\bibnamefont {Kunii}}, \bibinfo {author}
  {\bibfnamefont {P.-J.}\ \bibnamefont {Brown}}, \bibinfo {author}
  {\bibfnamefont {F.}~\bibnamefont {Tasset}}, \ and\ \bibinfo {author}
  {\bibfnamefont {T.}~\bibnamefont {Hansen}},\
  }\href{\doibase/10.1103/PhysRevB.68.214401} {\bibfield  {journal} {\bibinfo
  {journal} {Phys. Rev.~B}\ }\textbf {\bibinfo {volume} {68}},\ \bibinfo
  {pages} {214401} (\bibinfo {year} {2003})}\BibitemShut {NoStop}%
\bibitem [{\citenamefont {Kusunose}(2008)}]{Kusunose08}%
  \BibitemOpen
  \bibfield  {author} {\bibinfo {author} {\bibfnamefont {H.}~\bibnamefont
  {Kusunose}},\ }\href{https://doi.org/10.1143/jpsj.77.064710} {\bibfield
  {journal} {\bibinfo  {journal} {J.~Phys. Soc. Jpn.}\ }\textbf {\bibinfo
  {volume} {77}},\ \bibinfo {pages} {064710} (\bibinfo {year}
  {2008})}\BibitemShut {NoStop}%
\bibitem [{\citenamefont {Sakai}\ \emph {\textit{et~al.}}(1997)\citenamefont
  {Sakai}, \citenamefont {Shiina}, \citenamefont {Shiba},\ and\ \citenamefont
  {Thalmeier}}]{SakaiShiina97}%
  \BibitemOpen
  \bibfield  {author} {\bibinfo {author} {\bibfnamefont {O.}~\bibnamefont
  {Sakai}}, \bibinfo {author} {\bibfnamefont {R.}~\bibnamefont {Shiina}},
  \bibinfo {author} {\bibfnamefont {H.}~\bibnamefont {Shiba}}, \ and\ \bibinfo
  {author} {\bibfnamefont {P.}~\bibnamefont {Thalmeier}},\
  }\href{\doibase/10.1143/jpsj.66.3005} {\bibfield  {journal} {\bibinfo
  {journal} {J.~Phys. Soc. Jpn.}\ }\textbf {\bibinfo {volume} {66}},\ \bibinfo
  {pages} {3005} (\bibinfo {year} {1997})}\BibitemShut {NoStop}%
\bibitem [{\citenamefont {Shiina}\ \emph {\textit{et~al.}}(1997)\citenamefont
  {Shiina}, \citenamefont {Shiba},\ and\ \citenamefont
  {Thalmeier}}]{ShiinaShiba97}%
  \BibitemOpen
  \bibfield  {author} {\bibinfo {author} {\bibfnamefont {R.}~\bibnamefont
  {Shiina}}, \bibinfo {author} {\bibfnamefont {H.}~\bibnamefont {Shiba}}, \
  and\ \bibinfo {author} {\bibfnamefont {P.}~\bibnamefont {Thalmeier}},\
  }\href{https://doi.org/10.1143/jpsj.66.1741} {\bibfield  {journal} {\bibinfo
  {journal} {J.~Phys. Soc. Jpn.}\ }\textbf {\bibinfo {volume} {66}},\ \bibinfo
  {pages} {1741} (\bibinfo {year} {1997})}\BibitemShut {NoStop}%
\bibitem [{\citenamefont {Shiina}\ \emph {\textit{et~al.}}(1998)\citenamefont
  {Shiina}, \citenamefont {Sakai}, \citenamefont {Shiba},\ and\ \citenamefont
  {Thalmeier}}]{ShiinaSakai98}%
  \BibitemOpen
  \bibfield  {author} {\bibinfo {author} {\bibfnamefont {R.}~\bibnamefont
  {Shiina}}, \bibinfo {author} {\bibfnamefont {O.}~\bibnamefont {Sakai}},
  \bibinfo {author} {\bibfnamefont {H.}~\bibnamefont {Shiba}}, \ and\ \bibinfo
  {author} {\bibfnamefont {P.}~\bibnamefont {Thalmeier}},\
  }\href{\doibase/10.1143/JPSJ.67.941} {\bibfield  {journal} {\bibinfo
  {journal} {J.~Phys. Soc. Jpn.}\ }\textbf {\bibinfo {volume} {67}},\ \bibinfo
  {pages} {941} (\bibinfo {year} {1998})}\BibitemShut {NoStop}%
\bibitem [{\citenamefont {Sera}\ and\ \citenamefont
  {Kobayashi}(1999)}]{SeraKobayashi99}%
  \BibitemOpen
  \bibfield  {author} {\bibinfo {author} {\bibfnamefont {M.}~\bibnamefont
  {Sera}}\ and\ \bibinfo {author} {\bibfnamefont {S.}~\bibnamefont
  {Kobayashi}},\ }\href{\doibase/10.1143/JPSJ.68.1664} {\bibfield  {journal}
  {\bibinfo  {journal} {J.~Phys. Soc. Jpn.}\ }\textbf {\bibinfo {volume}
  {68}},\ \bibinfo {pages} {1664} (\bibinfo {year} {1999})}\BibitemShut
  {NoStop}%
\bibitem [{\citenamefont {Demishev}\ \emph {\textit{et~al.}}(2017)\citenamefont
  {Demishev}, \citenamefont {Krasnorussky}, \citenamefont {Bogach},
  \citenamefont {Voronov}, \citenamefont {Shitsevalova}, \citenamefont
  {Filipov}, \citenamefont {Glushkov},\ and\ \citenamefont
  {Sluchanko}}]{DemishevKrasnorussky17}%
  \BibitemOpen
  \bibfield  {author} {\bibinfo {author} {\bibfnamefont {S.~V.}\ \bibnamefont
  {Demishev}}, \bibinfo {author} {\bibfnamefont {V.~N.}\ \bibnamefont
  {Krasnorussky}}, \bibinfo {author} {\bibfnamefont {A.~V.}\ \bibnamefont
  {Bogach}}, \bibinfo {author} {\bibfnamefont {V.~V.}\ \bibnamefont {Voronov}},
  \bibinfo {author} {\bibfnamefont {N.~Y.}\ \bibnamefont {Shitsevalova}},
  \bibinfo {author} {\bibfnamefont {V.~B.}\ \bibnamefont {Filipov}}, \bibinfo
  {author} {\bibfnamefont {V.~V.}\ \bibnamefont {Glushkov}}, \ and\ \bibinfo
  {author} {\bibfnamefont {N.~E.}\ \bibnamefont {Sluchanko}},\
  }\href{https://doi.org/10.1038/s41598-017-17608-3} {\bibfield  {journal}
  {\bibinfo  {journal} {Sci. Rep.}\ }\textbf {\bibinfo {volume} {7}},\ \bibinfo
  {pages} {17430} (\bibinfo {year} {2017})}\BibitemShut {NoStop}%
\bibitem [{\citenamefont {Sera}\ \emph {\textit{et~al.}}(1987)\citenamefont
  {Sera}, \citenamefont {Sato},\ and\ \citenamefont {Kasuya}}]{SeraSato87}%
  \BibitemOpen
  \bibfield  {author} {\bibinfo {author} {\bibfnamefont {M.}~\bibnamefont
  {Sera}}, \bibinfo {author} {\bibfnamefont {N.}~\bibnamefont {Sato}}, \ and\
  \bibinfo {author} {\bibfnamefont {T.}~\bibnamefont {Kasuya}},\
  }\href{\doibase/10.1016/0304-8853(87)90523-3} {\bibfield  {journal} {\bibinfo
   {journal} {J. Magn. Magn. Mater.}\ }\textbf {\bibinfo {volume} {63}},\
  \bibinfo {pages} {64} (\bibinfo {year} {1987})}\BibitemShut {NoStop}%
\bibitem [{\citenamefont {Tayama}\ \emph {\textit{et~al.}}(1997)\citenamefont
  {Tayama}, \citenamefont {Sakakibara}, \citenamefont {Tenya}, \citenamefont
  {Amitsuka},\ and\ \citenamefont {Kunii}}]{TayamaSakakibara97}%
  \BibitemOpen
  \bibfield  {author} {\bibinfo {author} {\bibfnamefont {T.}~\bibnamefont
  {Tayama}}, \bibinfo {author} {\bibfnamefont {T.}~\bibnamefont {Sakakibara}},
  \bibinfo {author} {\bibfnamefont {K.}~\bibnamefont {Tenya}}, \bibinfo
  {author} {\bibfnamefont {H.}~\bibnamefont {Amitsuka}}, \ and\ \bibinfo
  {author} {\bibfnamefont {S.}~\bibnamefont {Kunii}},\
  }\href{https://doi.org/10.1143/jpsj.66.2268} {\bibfield  {journal} {\bibinfo
  {journal} {J.~Phys. Soc. Jpn.}\ }\textbf {\bibinfo {volume} {66}},\ \bibinfo
  {pages} {2268} (\bibinfo {year} {1997})}\BibitemShut {NoStop}%
\bibitem [{\citenamefont {i.~Kobayashi}\ \emph
  {\textit{et~al.}}(2000)\citenamefont {i.~Kobayashi}, \citenamefont {Sera},
  \citenamefont {Hiroi}, \citenamefont {Kobayashi},\ and\ \citenamefont
  {Kunii}}]{KobayashiSera00}%
  \BibitemOpen
  \bibfield  {author} {\bibinfo {author} {\bibfnamefont {S.}~\bibnamefont
  {i.~Kobayashi}}, \bibinfo {author} {\bibfnamefont {M.}~\bibnamefont {Sera}},
  \bibinfo {author} {\bibfnamefont {M.}~\bibnamefont {Hiroi}}, \bibinfo
  {author} {\bibfnamefont {N.}~\bibnamefont {Kobayashi}}, \ and\ \bibinfo
  {author} {\bibfnamefont {S.}~\bibnamefont {Kunii}},\
  }\href{https://doi.org/10.1016/s0921-4526(99)01223-5} {\bibfield  {journal}
  {\bibinfo  {journal} {Physica~B}\ }\textbf {\bibinfo {volume} {281--282}},\
  \bibinfo {pages} {557} (\bibinfo {year} {2000})}\BibitemShut {NoStop}%
\bibitem [{\citenamefont {Kobayashi}\ \emph
  {\textit{et~al.}}(2003)\citenamefont {Kobayashi}, \citenamefont {Yoshino},
  \citenamefont {Tsuji}, \citenamefont {Sera},\ and\ \citenamefont
  {Iga}}]{KobayashiYoshino03}%
  \BibitemOpen
  \bibfield  {author} {\bibinfo {author} {\bibfnamefont {S.}~\bibnamefont
  {Kobayashi}}, \bibinfo {author} {\bibfnamefont {Y.}~\bibnamefont {Yoshino}},
  \bibinfo {author} {\bibfnamefont {S.}~\bibnamefont {Tsuji}}, \bibinfo
  {author} {\bibfnamefont {M.}~\bibnamefont {Sera}}, \ and\ \bibinfo {author}
  {\bibfnamefont {F.}~\bibnamefont {Iga}},\ }\href{\doibase/10.1143/jpsj.72.25}
  {\bibfield  {journal} {\bibinfo  {journal} {J.~Phys. Soc. Jpn.}\ }\textbf
  {\bibinfo {volume} {72}},\ \bibinfo {pages} {25} (\bibinfo {year}
  {2003})}\BibitemShut {NoStop}%
\bibitem [{\citenamefont {Yoshino}\ \emph {\textit{et~al.}}(2004)\citenamefont
  {Yoshino}, \citenamefont {Kobayashi}, \citenamefont {Tsuji}, \citenamefont
  {Tou}, \citenamefont {Sera}, \citenamefont {Iga}, \citenamefont {Zenitani},\
  and\ \citenamefont {Akimitsu}}]{YoshinoKobayashi04}%
  \BibitemOpen
  \bibfield  {author} {\bibinfo {author} {\bibfnamefont {Y.}~\bibnamefont
  {Yoshino}}, \bibinfo {author} {\bibfnamefont {S.}~\bibnamefont {Kobayashi}},
  \bibinfo {author} {\bibfnamefont {S.}~\bibnamefont {Tsuji}}, \bibinfo
  {author} {\bibfnamefont {H.}~\bibnamefont {Tou}}, \bibinfo {author}
  {\bibfnamefont {M.}~\bibnamefont {Sera}}, \bibinfo {author} {\bibfnamefont
  {F.}~\bibnamefont {Iga}}, \bibinfo {author} {\bibfnamefont {Y.}~\bibnamefont
  {Zenitani}}, \ and\ \bibinfo {author} {\bibfnamefont {J.}~\bibnamefont
  {Akimitsu}},\ }\href{\doibase/10.1143/JPSJ.73.29} {\bibfield  {journal}
  {\bibinfo  {journal} {J.~Phys. Soc. Jpn.}\ }\textbf {\bibinfo {volume}
  {73}},\ \bibinfo {pages} {29} (\bibinfo {year} {2004})}\BibitemShut {NoStop}%
\bibitem [{\citenamefont {Mignot}\ \emph {\textit{et~al.}}(2009)\citenamefont
  {Mignot}, \citenamefont {Robert}, \citenamefont {Andr{\'e}}, \citenamefont
  {Sera},\ and\ \citenamefont {Iga}}]{MignotRobert09}%
  \BibitemOpen
  \bibfield  {author} {\bibinfo {author} {\bibfnamefont {J.~M.}\ \bibnamefont
  {Mignot}}, \bibinfo {author} {\bibfnamefont {J.}~\bibnamefont {Robert}},
  \bibinfo {author} {\bibfnamefont {G.}~\bibnamefont {Andr{\'e}}}, \bibinfo
  {author} {\bibfnamefont {M.}~\bibnamefont {Sera}}, \ and\ \bibinfo {author}
  {\bibfnamefont {F.}~\bibnamefont {Iga}},\
  }\href{\doibase/10.1103/PhysRevB.79.224426} {\bibfield  {journal} {\bibinfo
  {journal} {Phys. Rev.~B}\ }\textbf {\bibinfo {volume} {79}},\ \bibinfo
  {pages} {224426} (\bibinfo {year} {2009})}\BibitemShut {NoStop}%
\bibitem [{\citenamefont {Schenck}\ \emph {\textit{et~al.}}(2007)\citenamefont
  {Schenck}, \citenamefont {Gygax},\ and\ \citenamefont
  {Solt}}]{SchenckGygax07}%
  \BibitemOpen
  \bibfield  {author} {\bibinfo {author} {\bibfnamefont {A.}~\bibnamefont
  {Schenck}}, \bibinfo {author} {\bibfnamefont {F.~N.}\ \bibnamefont {Gygax}},
  \ and\ \bibinfo {author} {\bibfnamefont {G.}~\bibnamefont {Solt}},\
  }\href{https://link.aps.org/doi/10.1103/PhysRevB.75.024428} {\bibfield
  {journal} {\bibinfo  {journal} {Phys. Rev.~B}\ }\textbf {\bibinfo {volume}
  {75}},\ \bibinfo {pages} {024428} (\bibinfo {year} {2007})}\BibitemShut
  {NoStop}%
\bibitem [{\citenamefont {Kunimori}\ \emph {\textit{et~al.}}(2010)\citenamefont
  {Kunimori}, \citenamefont {Tanida}, \citenamefont {Matsumura}, \citenamefont
  {Sera},\ and\ \citenamefont {Iga}}]{KunimoriTanida10}%
  \BibitemOpen
  \bibfield  {author} {\bibinfo {author} {\bibfnamefont {K.}~\bibnamefont
  {Kunimori}}, \bibinfo {author} {\bibfnamefont {H.}~\bibnamefont {Tanida}},
  \bibinfo {author} {\bibfnamefont {T.}~\bibnamefont {Matsumura}}, \bibinfo
  {author} {\bibfnamefont {M.}~\bibnamefont {Sera}}, \ and\ \bibinfo {author}
  {\bibfnamefont {F.}~\bibnamefont {Iga}},\
  }\href{https://doi.org/10.1143/jpsj.79.073703} {\bibfield  {journal}
  {\bibinfo  {journal} {J.~Phys. Soc. Jpn.}\ }\textbf {\bibinfo {volume}
  {79}},\ \bibinfo {pages} {073703} (\bibinfo {year} {2010})}\BibitemShut
  {NoStop}%
\bibitem [{\citenamefont {Friemel}\ \emph {\textit{et~al.}}(2015)\citenamefont
  {Friemel}, \citenamefont {Jang}, \citenamefont {Schneidewind}, \citenamefont
  {Ivanov}, \citenamefont {Dukhnenko}, \citenamefont {Shitsevalova},
  \citenamefont {Filipov}, \citenamefont {Keimer},\ and\ \citenamefont
  {Inosov}}]{FriemelJang15}%
  \BibitemOpen
  \bibfield  {author} {\bibinfo {author} {\bibfnamefont {G.}~\bibnamefont
  {Friemel}}, \bibinfo {author} {\bibfnamefont {H.}~\bibnamefont {Jang}},
  \bibinfo {author} {\bibfnamefont {A.}~\bibnamefont {Schneidewind}}, \bibinfo
  {author} {\bibfnamefont {A.}~\bibnamefont {Ivanov}}, \bibinfo {author}
  {\bibfnamefont {A.~V.}\ \bibnamefont {Dukhnenko}}, \bibinfo {author}
  {\bibfnamefont {N.~Y.}\ \bibnamefont {Shitsevalova}}, \bibinfo {author}
  {\bibfnamefont {V.~B.}\ \bibnamefont {Filipov}}, \bibinfo {author}
  {\bibfnamefont {B.}~\bibnamefont {Keimer}}, \ and\ \bibinfo {author}
  {\bibfnamefont {D.~S.}\ \bibnamefont {Inosov}},\
  }\href{\doibase/10.1103/PhysRevB.92.014410} {\bibfield  {journal} {\bibinfo
  {journal} {Phys. Rev.~B}\ }\textbf {\bibinfo {volume} {92}},\ \bibinfo
  {pages} {014410} (\bibinfo {year} {2015})}\BibitemShut {NoStop}%
\bibitem [{\citenamefont {Matthias}\ \emph {\textit{et~al.}}(1968)\citenamefont
  {Matthias}, \citenamefont {Geballe}, \citenamefont {Andres}, \citenamefont
  {Corenzwit}, \citenamefont {Hull},\ and\ \citenamefont
  {Maita}}]{MatthiasGeballe68}%
  \BibitemOpen
  \bibfield  {author} {\bibinfo {author} {\bibfnamefont {B.~T.}\ \bibnamefont
  {Matthias}}, \bibinfo {author} {\bibfnamefont {T.~H.}\ \bibnamefont
  {Geballe}}, \bibinfo {author} {\bibfnamefont {K.}~\bibnamefont {Andres}},
  \bibinfo {author} {\bibfnamefont {E.}~\bibnamefont {Corenzwit}}, \bibinfo
  {author} {\bibfnamefont {G.~W.}\ \bibnamefont {Hull}}, \ and\ \bibinfo
  {author} {\bibfnamefont {J.~P.}\ \bibnamefont {Maita}},\
  }\href{http://science.sciencemag.org/content/159/3814/530} {\bibfield
  {journal} {\bibinfo  {journal} {Science}\ }\textbf {\bibinfo {volume}
  {159}},\ \bibinfo {pages} {530} (\bibinfo {year} {1968})}\BibitemShut
  {NoStop}%
\bibitem [{\citenamefont {Arko}\ \emph {\textit{et~al.}}(1975)\citenamefont
  {Arko}, \citenamefont {Crabtree}, \citenamefont {Ketterson}, \citenamefont
  {Mueller}, \citenamefont {Walch}, \citenamefont {Windmiller}, \citenamefont
  {Fisk}, \citenamefont {Hoyt}, \citenamefont {Mota}, \citenamefont
  {Viswanathan}, \citenamefont {Ellis}, \citenamefont {Freeman},\ and\
  \citenamefont {Rath}}]{ArkoCrabtree75}%
  \BibitemOpen
  \bibfield  {author} {\bibinfo {author} {\bibfnamefont {A.~J.}\ \bibnamefont
  {Arko}}, \bibinfo {author} {\bibfnamefont {G.}~\bibnamefont {Crabtree}},
  \bibinfo {author} {\bibfnamefont {J.~B.}\ \bibnamefont {Ketterson}}, \bibinfo
  {author} {\bibfnamefont {F.~M.}\ \bibnamefont {Mueller}}, \bibinfo {author}
  {\bibfnamefont {P.~F.}\ \bibnamefont {Walch}}, \bibinfo {author}
  {\bibfnamefont {L.~R.}\ \bibnamefont {Windmiller}}, \bibinfo {author}
  {\bibfnamefont {Z.}~\bibnamefont {Fisk}}, \bibinfo {author} {\bibfnamefont
  {R.~F.}\ \bibnamefont {Hoyt}}, \bibinfo {author} {\bibfnamefont {A.~C.}\
  \bibnamefont {Mota}}, \bibinfo {author} {\bibfnamefont {R.}~\bibnamefont
  {Viswanathan}}, \bibinfo {author} {\bibfnamefont {D.~E.}\ \bibnamefont
  {Ellis}}, \bibinfo {author} {\bibfnamefont {A.~J.}\ \bibnamefont {Freeman}},
  \ and\ \bibinfo {author} {\bibfnamefont {J.}~\bibnamefont {Rath}},\
  }\href{https://doi.org/10.1002/qua.560090868} {\bibfield  {journal} {\bibinfo
   {journal} {Int. J. Quant. Chem.}\ }\textbf {\bibinfo {volume} {9}},\
  \bibinfo {pages} {569} (\bibinfo {year} {1975})}\BibitemShut {NoStop}%
\bibitem [{\citenamefont {Bat'ko}\ \emph {\textit{et~al.}}(1995)\citenamefont
  {Bat'ko}, \citenamefont {Bat'kov{\'{a}}}, \citenamefont {Flachbart},
  \citenamefont {Filippov}, \citenamefont {Paderno}, \citenamefont
  {Shicevalova},\ and\ \citenamefont {Wagner}}]{BatkoBatkova95}%
  \BibitemOpen
  \bibfield  {author} {\bibinfo {author} {\bibfnamefont {I.}~\bibnamefont
  {Bat'ko}}, \bibinfo {author} {\bibfnamefont {M.}~\bibnamefont
  {Bat'kov{\'{a}}}}, \bibinfo {author} {\bibfnamefont {K.}~\bibnamefont
  {Flachbart}}, \bibinfo {author} {\bibfnamefont {V.~B.}\ \bibnamefont
  {Filippov}}, \bibinfo {author} {\bibfnamefont {Y.~B.}\ \bibnamefont
  {Paderno}}, \bibinfo {author} {\bibfnamefont {N.}~\bibnamefont
  {Shicevalova}}, \ and\ \bibinfo {author} {\bibfnamefont {T.}~\bibnamefont
  {Wagner}},\ }\href{https://doi.org/10.1016/0925-8388(94)01351-9} {\bibfield
  {journal} {\bibinfo  {journal} {J.~Alloy. Compd.}\ }\textbf {\bibinfo
  {volume} {217}},\ \bibinfo {pages} {L1} (\bibinfo {year} {1995})}\BibitemShut
  {NoStop}%
\bibitem [{\citenamefont {Friemel}\ \emph {\textit{et~al.}}(2012)\citenamefont
  {Friemel}, \citenamefont {Li}, \citenamefont {Dukhnenko}, \citenamefont
  {Shitsevalova}, \citenamefont {Sluchanko}, \citenamefont {Ivanov},
  \citenamefont {Filipov}, \citenamefont {Keimer},\ and\ \citenamefont
  {Inosov}}]{FriemelLi12}%
  \BibitemOpen
  \bibfield  {author} {\bibinfo {author} {\bibfnamefont {G.}~\bibnamefont
  {Friemel}}, \bibinfo {author} {\bibfnamefont {Y.}~\bibnamefont {Li}},
  \bibinfo {author} {\bibfnamefont {A.~V.}\ \bibnamefont {Dukhnenko}}, \bibinfo
  {author} {\bibfnamefont {N.~Y.}\ \bibnamefont {Shitsevalova}}, \bibinfo
  {author} {\bibfnamefont {N.~E.}\ \bibnamefont {Sluchanko}}, \bibinfo {author}
  {\bibfnamefont {A.}~\bibnamefont {Ivanov}}, \bibinfo {author} {\bibfnamefont
  {V.~B.}\ \bibnamefont {Filipov}}, \bibinfo {author} {\bibfnamefont
  {B.}~\bibnamefont {Keimer}}, \ and\ \bibinfo {author} {\bibfnamefont {D.~S.}\
  \bibnamefont {Inosov}},\ }\href{\doibase/10.1038/ncomms1821} {\bibfield
  {journal} {\bibinfo  {journal} {Nature Commun.}\ }\textbf {\bibinfo {volume}
  {3}},\ \bibinfo {pages} {830} (\bibinfo {year} {2012})}\BibitemShut {NoStop}%
\bibitem [{\citenamefont {Jang}\ \emph {\textit{et~al.}}(2014)\citenamefont
  {Jang}, \citenamefont {Friemel}, \citenamefont {Ollivier}, \citenamefont
  {Dukhnenko}, \citenamefont {Shitsevalova}, \citenamefont {Filipov},
  \citenamefont {Keimer},\ and\ \citenamefont {Inosov}}]{JangFriemel14}%
  \BibitemOpen
  \bibfield  {author} {\bibinfo {author} {\bibfnamefont {H.}~\bibnamefont
  {Jang}}, \bibinfo {author} {\bibfnamefont {G.}~\bibnamefont {Friemel}},
  \bibinfo {author} {\bibfnamefont {J.}~\bibnamefont {Ollivier}}, \bibinfo
  {author} {\bibfnamefont {A.~V.}\ \bibnamefont {Dukhnenko}}, \bibinfo {author}
  {\bibfnamefont {N.~Y.}\ \bibnamefont {Shitsevalova}}, \bibinfo {author}
  {\bibfnamefont {V.~B.}\ \bibnamefont {Filipov}}, \bibinfo {author}
  {\bibfnamefont {B.}~\bibnamefont {Keimer}}, \ and\ \bibinfo {author}
  {\bibfnamefont {D.~S.}\ \bibnamefont {Inosov}},\
  }\href{\doibase/10.1038/nmat3976} {\bibfield  {journal} {\bibinfo  {journal}
  {Nature Mater.}\ }\textbf {\bibinfo {volume} {13}},\ \bibinfo {pages} {682}
  (\bibinfo {year} {2014})}\BibitemShut {NoStop}%
\bibitem [{\citenamefont {{\=O}nuki}\ \emph
  {\textit{et~al.}}(1985)\citenamefont {{\=O}nuki}, \citenamefont {Shimizu},
  \citenamefont {Nishihara}, \citenamefont {Machii},\ and\ \citenamefont
  {Komatsubara}}]{OnukiShimizu85}%
  \BibitemOpen
  \bibfield  {author} {\bibinfo {author} {\bibfnamefont {Y.}~\bibnamefont
  {{\=O}nuki}}, \bibinfo {author} {\bibfnamefont {Y.}~\bibnamefont {Shimizu}},
  \bibinfo {author} {\bibfnamefont {M.}~\bibnamefont {Nishihara}}, \bibinfo
  {author} {\bibfnamefont {Y.}~\bibnamefont {Machii}}, \ and\ \bibinfo {author}
  {\bibfnamefont {T.}~\bibnamefont {Komatsubara}},\
  }\href{\doibase/10.1143/JPSJ.54.1964} {\bibfield  {journal} {\bibinfo
  {journal} {J.~Phys. Soc. Jpn.}\ }\textbf {\bibinfo {volume} {54}},\ \bibinfo
  {pages} {1964} (\bibinfo {year} {1985})}\BibitemShut {NoStop}%
\bibitem [{\citenamefont {Sumiyama}\ \emph {\textit{et~al.}}(1986)\citenamefont
  {Sumiyama}, \citenamefont {Oda}, \citenamefont {Nagano}, \citenamefont
  {{\={O}}nuki}, \citenamefont {Shibutani},\ and\ \citenamefont
  {Komatsubara}}]{SumiyamaOda86}%
  \BibitemOpen
  \bibfield  {author} {\bibinfo {author} {\bibfnamefont {A.}~\bibnamefont
  {Sumiyama}}, \bibinfo {author} {\bibfnamefont {Y.}~\bibnamefont {Oda}},
  \bibinfo {author} {\bibfnamefont {H.}~\bibnamefont {Nagano}}, \bibinfo
  {author} {\bibfnamefont {Y.}~\bibnamefont {{\={O}}nuki}}, \bibinfo {author}
  {\bibfnamefont {K.}~\bibnamefont {Shibutani}}, \ and\ \bibinfo {author}
  {\bibfnamefont {T.}~\bibnamefont {Komatsubara}},\
  }\href{https://doi.org/10.1143/jpsj.55.1294} {\bibfield  {journal} {\bibinfo
  {journal} {J.~Phys. Soc. Jpn.}\ }\textbf {\bibinfo {volume} {55}},\ \bibinfo
  {pages} {1294} (\bibinfo {year} {1986})}\BibitemShut {NoStop}%
\bibitem [{\citenamefont {Kato}\ \emph {\textit{et~al.}}(1987)\citenamefont
  {Kato}, \citenamefont {Satoh}, \citenamefont {Maeno}, \citenamefont {Aoki},
  \citenamefont {Fujita}, \citenamefont {{\={O}}nuki},\ and\ \citenamefont
  {Komatsubara}}]{KatoSatoh87}%
  \BibitemOpen
  \bibfield  {author} {\bibinfo {author} {\bibfnamefont {M.}~\bibnamefont
  {Kato}}, \bibinfo {author} {\bibfnamefont {K.}~\bibnamefont {Satoh}},
  \bibinfo {author} {\bibfnamefont {Y.}~\bibnamefont {Maeno}}, \bibinfo
  {author} {\bibfnamefont {Y.}~\bibnamefont {Aoki}}, \bibinfo {author}
  {\bibfnamefont {T.}~\bibnamefont {Fujita}}, \bibinfo {author} {\bibfnamefont
  {Y.}~\bibnamefont {{\={O}}nuki}}, \ and\ \bibinfo {author} {\bibfnamefont
  {T.}~\bibnamefont {Komatsubara}},\
  }\href{https://doi.org/10.1143/jpsj.56.3661} {\bibfield  {journal} {\bibinfo
  {journal} {J.~Phys. Soc. Jpn.}\ }\textbf {\bibinfo {volume} {56}},\ \bibinfo
  {pages} {3661} (\bibinfo {year} {1987})}\BibitemShut {NoStop}%
\bibitem [{\citenamefont {Satoh}\ \emph {\textit{et~al.}}(1989)\citenamefont
  {Satoh}, \citenamefont {Fujita}, \citenamefont {Maeno}, \citenamefont
  {{\={O}}nuki},\ and\ \citenamefont {Komatsubara}}]{SatohFujita89}%
  \BibitemOpen
  \bibfield  {author} {\bibinfo {author} {\bibfnamefont {K.}~\bibnamefont
  {Satoh}}, \bibinfo {author} {\bibfnamefont {T.}~\bibnamefont {Fujita}},
  \bibinfo {author} {\bibfnamefont {Y.}~\bibnamefont {Maeno}}, \bibinfo
  {author} {\bibfnamefont {Y.}~\bibnamefont {{\={O}}nuki}}, \ and\ \bibinfo
  {author} {\bibfnamefont {T.}~\bibnamefont {Komatsubara}},\
  }\href{https://doi.org/10.1143/jpsj.58.1012} {\bibfield  {journal} {\bibinfo
  {journal} {J.~Phys. Soc. Jpn.}\ }\textbf {\bibinfo {volume} {58}},\ \bibinfo
  {pages} {1012} (\bibinfo {year} {1989})}\BibitemShut {NoStop}%
\bibitem [{\citenamefont {Lovesey}\ \emph {\textit{et~al.}}(2007)\citenamefont
  {Lovesey}, \citenamefont {Fern\'andez-Rodr\'{\i}guez}, \citenamefont
  {Blanco},\ and\ \citenamefont {Tanaka}}]{LoveseyFernandezRodriguez07}%
  \BibitemOpen
  \bibfield  {author} {\bibinfo {author} {\bibfnamefont {S.~W.}\ \bibnamefont
  {Lovesey}}, \bibinfo {author} {\bibfnamefont {J.}~\bibnamefont
  {Fern\'andez-Rodr\'{\i}guez}}, \bibinfo {author} {\bibfnamefont {J.~A.}\
  \bibnamefont {Blanco}}, \ and\ \bibinfo {author} {\bibfnamefont
  {Y.}~\bibnamefont {Tanaka}},\
  }\href{https://link.aps.org/doi/10.1103/PhysRevB.75.054401} {\bibfield
  {journal} {\bibinfo  {journal} {Phys. Rev. B}\ }\textbf {\bibinfo {volume}
  {75}},\ \bibinfo {pages} {054401} (\bibinfo {year} {2007})}\BibitemShut
  {NoStop}%
\bibitem [{\citenamefont {Kuwahara}\ \emph {\textit{et~al.}}(2007)\citenamefont
  {Kuwahara}, \citenamefont {Iwasa}, \citenamefont {Kohgi}, \citenamefont
  {Aso}, \citenamefont {Sera},\ and\ \citenamefont {Iga}}]{KuwaharaIwasa07}%
  \BibitemOpen
  \bibfield  {author} {\bibinfo {author} {\bibfnamefont {K.}~\bibnamefont
  {Kuwahara}}, \bibinfo {author} {\bibfnamefont {K.}~\bibnamefont {Iwasa}},
  \bibinfo {author} {\bibfnamefont {M.}~\bibnamefont {Kohgi}}, \bibinfo
  {author} {\bibfnamefont {N.}~\bibnamefont {Aso}}, \bibinfo {author}
  {\bibfnamefont {M.}~\bibnamefont {Sera}}, \ and\ \bibinfo {author}
  {\bibfnamefont {F.}~\bibnamefont {Iga}},\
  }\href{https://doi.org/10.1143/jpsj.76.093702} {\bibfield  {journal}
  {\bibinfo  {journal} {J.~Phys. Soc. Jpn.}\ }\textbf {\bibinfo {volume}
  {76}},\ \bibinfo {pages} {093702} (\bibinfo {year} {2007})}\BibitemShut
  {NoStop}%
\bibitem [{\citenamefont {Kuwahara}\ \emph {\textit{et~al.}}(2009)\citenamefont
  {Kuwahara}, \citenamefont {Iwasa}, \citenamefont {Kohgi}, \citenamefont
  {Aso}, \citenamefont {Sera}, \citenamefont {Iga}, \citenamefont {Matsuura},\
  and\ \citenamefont {Hirota}}]{KuwaharaIwasa09}%
  \BibitemOpen
  \bibfield  {author} {\bibinfo {author} {\bibfnamefont {K.}~\bibnamefont
  {Kuwahara}}, \bibinfo {author} {\bibfnamefont {K.}~\bibnamefont {Iwasa}},
  \bibinfo {author} {\bibfnamefont {M.}~\bibnamefont {Kohgi}}, \bibinfo
  {author} {\bibfnamefont {N.}~\bibnamefont {Aso}}, \bibinfo {author}
  {\bibfnamefont {M.}~\bibnamefont {Sera}}, \bibinfo {author} {\bibfnamefont
  {F.}~\bibnamefont {Iga}}, \bibinfo {author} {\bibfnamefont {M.}~\bibnamefont
  {Matsuura}}, \ and\ \bibinfo {author} {\bibfnamefont {K.}~\bibnamefont
  {Hirota}},\ }\href{https://doi.org/10.1016/j.physb.2009.06.016} {\bibfield
  {journal} {\bibinfo  {journal} {Physica~B}\ }\textbf {\bibinfo {volume}
  {404}},\ \bibinfo {pages} {2527} (\bibinfo {year} {2009})}\BibitemShut
  {NoStop}%
\bibitem [{\citenamefont {Nakamura}\ \emph {\textit{et~al.}}(2006)\citenamefont
  {Nakamura}, \citenamefont {Endo}, \citenamefont {Yamamoto}, \citenamefont
  {Isshiki}, \citenamefont {Kimura}, \citenamefont {Aoki}, \citenamefont
  {Nojima}, \citenamefont {Otani},\ and\ \citenamefont
  {Kunii}}]{NakamuraEndo06}%
  \BibitemOpen
  \bibfield  {author} {\bibinfo {author} {\bibfnamefont {S.}~\bibnamefont
  {Nakamura}}, \bibinfo {author} {\bibfnamefont {M.}~\bibnamefont {Endo}},
  \bibinfo {author} {\bibfnamefont {H.}~\bibnamefont {Yamamoto}}, \bibinfo
  {author} {\bibfnamefont {T.}~\bibnamefont {Isshiki}}, \bibinfo {author}
  {\bibfnamefont {N.}~\bibnamefont {Kimura}}, \bibinfo {author} {\bibfnamefont
  {H.}~\bibnamefont {Aoki}}, \bibinfo {author} {\bibfnamefont {T.}~\bibnamefont
  {Nojima}}, \bibinfo {author} {\bibfnamefont {S.}~\bibnamefont {Otani}}, \
  and\ \bibinfo {author} {\bibfnamefont {S.}~\bibnamefont {Kunii}},\
  }\href{\doibase/10.1103/PhysRevLett.97.237204} {\bibfield  {journal}
  {\bibinfo  {journal} {Phys. Rev. Lett.}\ }\textbf {\bibinfo {volume} {97}},\
  \bibinfo {pages} {237204} (\bibinfo {year} {2006})}\BibitemShut {NoStop}%
\bibitem [{\citenamefont {Jang}\ \emph {\textit{et~al.}}(2017)\citenamefont
  {Jang}, \citenamefont {Portnichenko}, \citenamefont {Cameron}, \citenamefont
  {Friemel}, \citenamefont {Dukhnenko}, \citenamefont {Shitsevalova},
  \citenamefont {Filipov}, \citenamefont {Schneidewind}, \citenamefont
  {Ivanov}, \citenamefont {Inosov},\ and\ \citenamefont
  {Brando}}]{JangPortnichenko17}%
  \BibitemOpen
  \bibfield  {author} {\bibinfo {author} {\bibfnamefont {D.~J.}\ \bibnamefont
  {Jang}}, \bibinfo {author} {\bibfnamefont {P.~Y.}\ \bibnamefont
  {Portnichenko}}, \bibinfo {author} {\bibfnamefont {A.~S.}\ \bibnamefont
  {Cameron}}, \bibinfo {author} {\bibfnamefont {G.}~\bibnamefont {Friemel}},
  \bibinfo {author} {\bibfnamefont {A.~V.}\ \bibnamefont {Dukhnenko}}, \bibinfo
  {author} {\bibfnamefont {N.~Y.}\ \bibnamefont {Shitsevalova}}, \bibinfo
  {author} {\bibfnamefont {V.~B.}\ \bibnamefont {Filipov}}, \bibinfo {author}
  {\bibfnamefont {A.}~\bibnamefont {Schneidewind}}, \bibinfo {author}
  {\bibfnamefont {A.}~\bibnamefont {Ivanov}}, \bibinfo {author} {\bibfnamefont
  {D.~S.}\ \bibnamefont {Inosov}}, \ and\ \bibinfo {author} {\bibfnamefont
  {M.}~\bibnamefont {Brando}},\ }\href{\doibase/10.1038/s41535-017-0068-x}
  {\bibfield  {journal} {\bibinfo  {journal} {npj~Quant. Mater.}\ }\textbf
  {\bibinfo {volume} {2}},\ \bibinfo {pages} {62} (\bibinfo {year}
  {2017})}\BibitemShut {NoStop}%
\bibitem [{\citenamefont {Shiba}\ \emph {\textit{et~al.}}(1999)\citenamefont
  {Shiba}, \citenamefont {Sakai},\ and\ \citenamefont {Shiina}}]{ShibaSakai99}%
  \BibitemOpen
  \bibfield  {author} {\bibinfo {author} {\bibfnamefont {H.}~\bibnamefont
  {Shiba}}, \bibinfo {author} {\bibfnamefont {O.}~\bibnamefont {Sakai}}, \ and\
  \bibinfo {author} {\bibfnamefont {R.}~\bibnamefont {Shiina}},\
  }\href{https://doi.org/10.1143/jpsj.68.1988} {\bibfield  {journal} {\bibinfo
  {journal} {J.~Phys. Soc. Jpn.}\ }\textbf {\bibinfo {volume} {68}},\ \bibinfo
  {pages} {1988} (\bibinfo {year} {1999})}\BibitemShut {NoStop}%
\bibitem [{\citenamefont {Shiba}\ \emph {\textit{et~al.}}(2000)\citenamefont
  {Shiba}, \citenamefont {Sakai},\ and\ \citenamefont {Shiina}}]{ShibaSakai00}%
  \BibitemOpen
  \bibfield  {author} {\bibinfo {author} {\bibfnamefont {H.}~\bibnamefont
  {Shiba}}, \bibinfo {author} {\bibfnamefont {O.}~\bibnamefont {Sakai}}, \ and\
  \bibinfo {author} {\bibfnamefont {R.}~\bibnamefont {Shiina}},\
  }\href{http://www.sciencedirect.com/science/article/pii/S0921452699007905}
  {\bibfield  {journal} {\bibinfo  {journal} {Physica~B}\ }\textbf {\bibinfo
  {volume} {281--282}},\ \bibinfo {pages} {477} (\bibinfo {year}
  {2000})}\BibitemShut {NoStop}%
\bibitem [{\citenamefont {Hanzawa}\ and\ \citenamefont
  {Takasaki}(2002)}]{HanzawaTakasaki02}%
  \BibitemOpen
  \bibfield  {author} {\bibinfo {author} {\bibfnamefont {K.}~\bibnamefont
  {Hanzawa}}\ and\ \bibinfo {author} {\bibfnamefont {K.}~\bibnamefont
  {Takasaki}},\ }\href{https://doi.org/10.1143/jpsj.71.2953} {\bibfield
  {journal} {\bibinfo  {journal} {J.~Phys. Soc. Jpn.}\ }\textbf {\bibinfo
  {volume} {71}},\ \bibinfo {pages} {2953} (\bibinfo {year}
  {2002})}\BibitemShut {NoStop}%
\bibitem [{\citenamefont {Hanzawa}(2004)}]{Hanzawa04}%
  \BibitemOpen
  \bibfield  {author} {\bibinfo {author} {\bibfnamefont {K.}~\bibnamefont
  {Hanzawa}},\ }\href{https://doi.org/10.1143/jpsj.73.1228} {\bibfield
  {journal} {\bibinfo  {journal} {J.~Phys. Soc. Jpn.}\ }\textbf {\bibinfo
  {volume} {73}},\ \bibinfo {pages} {1228} (\bibinfo {year}
  {2004})}\BibitemShut {NoStop}%
\bibitem [{\citenamefont {Sakurai}\ and\ \citenamefont
  {Kuramoto}(2005)}]{SakuraiKuramoto05}%
  \BibitemOpen
  \bibfield  {author} {\bibinfo {author} {\bibfnamefont {G.}~\bibnamefont
  {Sakurai}}\ and\ \bibinfo {author} {\bibfnamefont {Y.}~\bibnamefont
  {Kuramoto}},\ }\href{https://doi.org/10.1143/jpsj.74.975} {\bibfield
  {journal} {\bibinfo  {journal} {J.~Phys. Soc. Jpn.}\ }\textbf {\bibinfo
  {volume} {74}},\ \bibinfo {pages} {975} (\bibinfo {year} {2005})}\BibitemShut
  {NoStop}%
\bibitem [{\citenamefont {Kondo}\ \emph {\textit{et~al.}}(2007)\citenamefont
  {Kondo}, \citenamefont {Tou}, \citenamefont {Sera}, \citenamefont {Iga},\
  and\ \citenamefont {Sakakibara}}]{KondoTou07}%
  \BibitemOpen
  \bibfield  {author} {\bibinfo {author} {\bibfnamefont {A.}~\bibnamefont
  {Kondo}}, \bibinfo {author} {\bibfnamefont {H.}~\bibnamefont {Tou}}, \bibinfo
  {author} {\bibfnamefont {M.}~\bibnamefont {Sera}}, \bibinfo {author}
  {\bibfnamefont {F.}~\bibnamefont {Iga}}, \ and\ \bibinfo {author}
  {\bibfnamefont {T.}~\bibnamefont {Sakakibara}},\
  }\href{https://doi.org/10.1143/jpsj.76.103708} {\bibfield  {journal}
  {\bibinfo  {journal} {J.~Phys. Soc. Jpn.}\ }\textbf {\bibinfo {volume}
  {76}},\ \bibinfo {pages} {103708} (\bibinfo {year} {2007})}\BibitemShut
  {NoStop}%
\bibitem [{\citenamefont {Kishimoto}\ \emph
  {\textit{et~al.}}(2005)\citenamefont {Kishimoto}, \citenamefont {Kondo},
  \citenamefont {Kim}, \citenamefont {Tou}, \citenamefont {Sera},\ and\
  \citenamefont {Iga}}]{KishimotoKondo05}%
  \BibitemOpen
  \bibfield  {author} {\bibinfo {author} {\bibfnamefont {S.}~\bibnamefont
  {Kishimoto}}, \bibinfo {author} {\bibfnamefont {A.}~\bibnamefont {Kondo}},
  \bibinfo {author} {\bibfnamefont {M.-S.}\ \bibnamefont {Kim}}, \bibinfo
  {author} {\bibfnamefont {H.}~\bibnamefont {Tou}}, \bibinfo {author}
  {\bibfnamefont {M.}~\bibnamefont {Sera}}, \ and\ \bibinfo {author}
  {\bibfnamefont {F.}~\bibnamefont {Iga}},\
  }\href{\doibase/10.1143/jpsj.74.2913} {\bibfield  {journal} {\bibinfo
  {journal} {J.~Phys. Soc. Jpn.}\ }\textbf {\bibinfo {volume} {74}},\ \bibinfo
  {pages} {2913} (\bibinfo {year} {2005})}\BibitemShut {NoStop}%
\bibitem [{\citenamefont {Matsumura}\ \emph
  {\textit{et~al.}}(2014{\natexlab{a}})\citenamefont {Matsumura}, \citenamefont
  {Kunimori}, \citenamefont {Kondo}, \citenamefont {Soejima}, \citenamefont
  {Tanida}, \citenamefont {Mignot}, \citenamefont {Iga},\ and\ \citenamefont
  {Sera}}]{MatsumuraKunimori14}%
  \BibitemOpen
  \bibfield  {author} {\bibinfo {author} {\bibfnamefont {T.}~\bibnamefont
  {Matsumura}}, \bibinfo {author} {\bibfnamefont {K.}~\bibnamefont {Kunimori}},
  \bibinfo {author} {\bibfnamefont {A.}~\bibnamefont {Kondo}}, \bibinfo
  {author} {\bibfnamefont {K.}~\bibnamefont {Soejima}}, \bibinfo {author}
  {\bibfnamefont {H.}~\bibnamefont {Tanida}}, \bibinfo {author} {\bibfnamefont
  {J.-M.}\ \bibnamefont {Mignot}}, \bibinfo {author} {\bibfnamefont
  {F.}~\bibnamefont {Iga}}, \ and\ \bibinfo {author} {\bibfnamefont
  {M.}~\bibnamefont {Sera}},\ }\href{https://doi.org/10.7566/jpsj.83.094724}
  {\bibfield  {journal} {\bibinfo  {journal} {J.~Phys. Soc. Jpn.}\ }\textbf
  {\bibinfo {volume} {83}},\ \bibinfo {pages} {094724} (\bibinfo {year}
  {2014}{\natexlab{a}})}\BibitemShut {NoStop}%
\bibitem [{\citenamefont {Onuki}\ \emph {\textit{et~al.}}(1989)\citenamefont
  {Onuki}, \citenamefont {Umezawa}, \citenamefont {Kwok}, \citenamefont
  {Crabtree}, \citenamefont {Nishihara}, \citenamefont {Yamazaki},
  \citenamefont {Omi},\ and\ \citenamefont {Komatsubara}}]{OnukiUmezawa89}%
  \BibitemOpen
  \bibfield  {author} {\bibinfo {author} {\bibfnamefont {Y.}~\bibnamefont
  {Onuki}}, \bibinfo {author} {\bibfnamefont {A.}~\bibnamefont {Umezawa}},
  \bibinfo {author} {\bibfnamefont {W.~K.}\ \bibnamefont {Kwok}}, \bibinfo
  {author} {\bibfnamefont {G.~W.}\ \bibnamefont {Crabtree}}, \bibinfo {author}
  {\bibfnamefont {M.}~\bibnamefont {Nishihara}}, \bibinfo {author}
  {\bibfnamefont {T.}~\bibnamefont {Yamazaki}}, \bibinfo {author}
  {\bibfnamefont {T.}~\bibnamefont {Omi}}, \ and\ \bibinfo {author}
  {\bibfnamefont {T.}~\bibnamefont {Komatsubara}},\
  }\href{\doibase/10.1103/PhysRevB.40.11195} {\bibfield  {journal} {\bibinfo
  {journal} {Phys. Rev.~B}\ }\textbf {\bibinfo {volume} {40}},\ \bibinfo
  {pages} {11195} (\bibinfo {year} {1989})}\BibitemShut {NoStop}%
\bibitem [{\citenamefont {Arko}\ \emph {\textit{et~al.}}(1976)\citenamefont
  {Arko}, \citenamefont {Crabtree}, \citenamefont {Karim}, \citenamefont
  {Mueller}, \citenamefont {Windmiller}, \citenamefont {Ketterson},\ and\
  \citenamefont {Fisk}}]{ArkoCrabtree76}%
  \BibitemOpen
  \bibfield  {author} {\bibinfo {author} {\bibfnamefont {A.~J.}\ \bibnamefont
  {Arko}}, \bibinfo {author} {\bibfnamefont {G.}~\bibnamefont {Crabtree}},
  \bibinfo {author} {\bibfnamefont {D.}~\bibnamefont {Karim}}, \bibinfo
  {author} {\bibfnamefont {F.~M.}\ \bibnamefont {Mueller}}, \bibinfo {author}
  {\bibfnamefont {L.~R.}\ \bibnamefont {Windmiller}}, \bibinfo {author}
  {\bibfnamefont {J.~B.}\ \bibnamefont {Ketterson}}, \ and\ \bibinfo {author}
  {\bibfnamefont {Z.}~\bibnamefont {Fisk}},\
  }\href{\doibase/10.1103/PhysRevB.13.5240} {\bibfield  {journal} {\bibinfo
  {journal} {Phys. Rev.~B}\ }\textbf {\bibinfo {volume} {13}},\ \bibinfo
  {pages} {5240} (\bibinfo {year} {1976})}\BibitemShut {NoStop}%
\bibitem [{\citenamefont {Neupane}\ \emph {\textit{et~al.}}(2015)\citenamefont
  {Neupane}, \citenamefont {Alidoust}, \citenamefont {Belopolski},
  \citenamefont {Bian}, \citenamefont {Xu}, \citenamefont {Kim}, \citenamefont
  {Shibayev}, \citenamefont {Sanchez}, \citenamefont {Zheng}, \citenamefont
  {Chang}, \citenamefont {Jeng}, \citenamefont {Riseborough}, \citenamefont
  {Lin}, \citenamefont {Bansil}, \citenamefont {Durakiewicz}, \citenamefont
  {Fisk},\ and\ \citenamefont {Hasan}}]{NeupaneAlidoust15}%
  \BibitemOpen
  \bibfield  {author} {\bibinfo {author} {\bibfnamefont {M.}~\bibnamefont
  {Neupane}}, \bibinfo {author} {\bibfnamefont {N.}~\bibnamefont {Alidoust}},
  \bibinfo {author} {\bibfnamefont {I.}~\bibnamefont {Belopolski}}, \bibinfo
  {author} {\bibfnamefont {G.}~\bibnamefont {Bian}}, \bibinfo {author}
  {\bibfnamefont {S.-Y.}\ \bibnamefont {Xu}}, \bibinfo {author} {\bibfnamefont
  {D.-J.}\ \bibnamefont {Kim}}, \bibinfo {author} {\bibfnamefont {P.~P.}\
  \bibnamefont {Shibayev}}, \bibinfo {author} {\bibfnamefont {D.~S.}\
  \bibnamefont {Sanchez}}, \bibinfo {author} {\bibfnamefont {H.}~\bibnamefont
  {Zheng}}, \bibinfo {author} {\bibfnamefont {T.-R.}\ \bibnamefont {Chang}},
  \bibinfo {author} {\bibfnamefont {H.-T.}\ \bibnamefont {Jeng}}, \bibinfo
  {author} {\bibfnamefont {P.~S.}\ \bibnamefont {Riseborough}}, \bibinfo
  {author} {\bibfnamefont {H.}~\bibnamefont {Lin}}, \bibinfo {author}
  {\bibfnamefont {A.}~\bibnamefont {Bansil}}, \bibinfo {author} {\bibfnamefont
  {T.}~\bibnamefont {Durakiewicz}}, \bibinfo {author} {\bibfnamefont
  {Z.}~\bibnamefont {Fisk}}, \ and\ \bibinfo {author} {\bibfnamefont {M.~Z.}\
  \bibnamefont {Hasan}},\ }\href{\doibase/10.1103/PhysRevB.92.104420}
  {\bibfield  {journal} {\bibinfo  {journal} {Phys. Rev.~B}\ }\textbf {\bibinfo
  {volume} {92}},\ \bibinfo {pages} {104420} (\bibinfo {year}
  {2015})}\BibitemShut {NoStop}%
\bibitem [{\citenamefont {Ollivier}\ and\ \citenamefont
  {Mutka}(2011)}]{OllivierMutka11}%
  \BibitemOpen
  \bibfield  {author} {\bibinfo {author} {\bibfnamefont {J.}~\bibnamefont
  {Ollivier}}\ and\ \bibinfo {author} {\bibfnamefont {H.}~\bibnamefont
  {Mutka}},\ }\href{\doibase/10.1143/JPSJS.80SB.SB003} {\bibfield  {journal}
  {\bibinfo  {journal} {J.~Phys. Soc. Jpn.}\ }\textbf {\bibinfo {volume}
  {80}},\ \bibinfo {pages} {SB003} (\bibinfo {year} {2011})}\BibitemShut
  {NoStop}%
\bibitem [{\citenamefont {Rodriguez}\ \emph
  {\textit{et~al.}}(2008)\citenamefont {Rodriguez}, \citenamefont {Adler},
  \citenamefont {Brand}, \citenamefont {Broholm}, \citenamefont {Cook},
  \citenamefont {Brocker}, \citenamefont {Hammond}, \citenamefont {Huang},
  \citenamefont {Hundertmark}, \citenamefont {Lynn}, \citenamefont
  {Maliszewskyj}, \citenamefont {Moyer}, \citenamefont {Orndorff},
  \citenamefont {Pierce}, \citenamefont {Pike}, \citenamefont {Scharfstein},
  \citenamefont {Smee},\ and\ \citenamefont {Vilaseca}}]{RodriguezAdler08}%
  \BibitemOpen
  \bibfield  {author} {\bibinfo {author} {\bibfnamefont {J.~A.}\ \bibnamefont
  {Rodriguez}}, \bibinfo {author} {\bibfnamefont {D.~M.}\ \bibnamefont
  {Adler}}, \bibinfo {author} {\bibfnamefont {P.~C.}\ \bibnamefont {Brand}},
  \bibinfo {author} {\bibfnamefont {C.}~\bibnamefont {Broholm}}, \bibinfo
  {author} {\bibfnamefont {J.~C.}\ \bibnamefont {Cook}}, \bibinfo {author}
  {\bibfnamefont {C.}~\bibnamefont {Brocker}}, \bibinfo {author} {\bibfnamefont
  {R.}~\bibnamefont {Hammond}}, \bibinfo {author} {\bibfnamefont
  {Z.}~\bibnamefont {Huang}}, \bibinfo {author} {\bibfnamefont
  {P.}~\bibnamefont {Hundertmark}}, \bibinfo {author} {\bibfnamefont {J.~W.}\
  \bibnamefont {Lynn}}, \bibinfo {author} {\bibfnamefont {N.}~\bibnamefont
  {Maliszewskyj}}, \bibinfo {author} {\bibfnamefont {J.}~\bibnamefont {Moyer}},
  \bibinfo {author} {\bibfnamefont {J.}~\bibnamefont {Orndorff}}, \bibinfo
  {author} {\bibfnamefont {D.}~\bibnamefont {Pierce}}, \bibinfo {author}
  {\bibfnamefont {T.~D.}\ \bibnamefont {Pike}}, \bibinfo {author}
  {\bibfnamefont {G.}~\bibnamefont {Scharfstein}}, \bibinfo {author}
  {\bibfnamefont {S.~A.}\ \bibnamefont {Smee}}, \ and\ \bibinfo {author}
  {\bibfnamefont {R.}~\bibnamefont {Vilaseca}},\
  }\href{\doibase/10.1088/0957-0233/19/3/034023} {\bibfield  {journal}
  {\bibinfo  {journal} {Meas. Sci. Technol.}\ }\textbf {\bibinfo {volume}
  {19}},\ \bibinfo {pages} {034023} (\bibinfo {year} {2008})}\BibitemShut
  {NoStop}%
\bibitem [{\citenamefont {Azuah}\ \emph {\textit{et~al.}}(2009)\citenamefont
  {Azuah}, \citenamefont {Kneller}, \citenamefont {Qiu}, \citenamefont
  {Tregenna-Piggott}, \citenamefont {Brown}, \citenamefont {Copley},\ and\
  \citenamefont {Dimeo}}]{AzuahKneller09}%
  \BibitemOpen
  \bibfield  {author} {\bibinfo {author} {\bibfnamefont {R.~T.}\ \bibnamefont
  {Azuah}}, \bibinfo {author} {\bibfnamefont {L.~R.}\ \bibnamefont {Kneller}},
  \bibinfo {author} {\bibfnamefont {Y.}~\bibnamefont {Qiu}}, \bibinfo {author}
  {\bibfnamefont {P.~L.~W.}\ \bibnamefont {Tregenna-Piggott}}, \bibinfo
  {author} {\bibfnamefont {C.~M.}\ \bibnamefont {Brown}}, \bibinfo {author}
  {\bibfnamefont {J.~R.~D.}\ \bibnamefont {Copley}}, \ and\ \bibinfo {author}
  {\bibfnamefont {R.~M.}\ \bibnamefont {Dimeo}},\
  }\href{\doibase/10.6028/jres.114.025} {\bibfield  {journal} {\bibinfo
  {journal} {J. Res. Natl. Inst. Stan. Technol.}\ }\textbf {\bibinfo {volume}
  {114}},\ \bibinfo {pages} {341} (\bibinfo {year} {2009})}\BibitemShut
  {NoStop}%
\bibitem [{\citenamefont {Ewings}\ \emph {\textit{et~al.}}(2016)\citenamefont
  {Ewings}, \citenamefont {Buts}, \citenamefont {Le}, \citenamefont {van
  Duijn}, \citenamefont {Bustinduy},\ and\ \citenamefont
  {Perring}}]{EwingsButs16}%
  \BibitemOpen
  \bibfield  {author} {\bibinfo {author} {\bibfnamefont {R.~A.}\ \bibnamefont
  {Ewings}}, \bibinfo {author} {\bibfnamefont {A.}~\bibnamefont {Buts}},
  \bibinfo {author} {\bibfnamefont {M.~D.}\ \bibnamefont {Le}}, \bibinfo
  {author} {\bibfnamefont {J.}~\bibnamefont {van Duijn}}, \bibinfo {author}
  {\bibfnamefont {I.}~\bibnamefont {Bustinduy}}, \ and\ \bibinfo {author}
  {\bibfnamefont {T.~G.}\ \bibnamefont {Perring}},\
  }\href{\doibase/10.1016/j.nima.2016.07.036} {\bibfield  {journal} {\bibinfo
  {journal} {Nucl. Instrum. Methods Phys. Res., Sect. A}\ }\textbf {\bibinfo
  {volume} {834}},\ \bibinfo {pages} {132} (\bibinfo {year}
  {2016})}\BibitemShut {NoStop}%
\bibitem [{\citenamefont {Tomita}\ and\ \citenamefont
  {Kunii}(1992)}]{TomitaKunii92}%
  \BibitemOpen
  \bibfield  {author} {\bibinfo {author} {\bibfnamefont {A.}~\bibnamefont
  {Tomita}}\ and\ \bibinfo {author} {\bibfnamefont {S.}~\bibnamefont {Kunii}},\
  }\href{\doibase/10.1016/0304-8853(92)91396-B} {\bibfield  {journal} {\bibinfo
   {journal} {J. Magn. Magn. Mater.}\ }\textbf {\bibinfo {volume} {108}},\
  \bibinfo {pages} {165} (\bibinfo {year} {1992})}\BibitemShut {NoStop}%
\bibitem [{\citenamefont {Anisimov}\ \emph {\textit{et~al.}}(2013)\citenamefont
  {Anisimov}, \citenamefont {Glushkov}, \citenamefont {Bogach}, \citenamefont
  {Demishev}, \citenamefont {Samarin}, \citenamefont {Gavrilkin}, \citenamefont
  {Mitsen}, \citenamefont {Shitsevalova}, \citenamefont {Levchenko},
  \citenamefont {Filippov}, \citenamefont {Gabani}, \citenamefont {Flachbart},\
  and\ \citenamefont {Sluchanko}}]{AnisimovGlushkov13}%
  \BibitemOpen
  \bibfield  {author} {\bibinfo {author} {\bibfnamefont {M.~A.}\ \bibnamefont
  {Anisimov}}, \bibinfo {author} {\bibfnamefont {V.~V.}\ \bibnamefont
  {Glushkov}}, \bibinfo {author} {\bibfnamefont {A.~V.}\ \bibnamefont
  {Bogach}}, \bibinfo {author} {\bibfnamefont {S.~V.}\ \bibnamefont
  {Demishev}}, \bibinfo {author} {\bibfnamefont {N.~A.}\ \bibnamefont
  {Samarin}}, \bibinfo {author} {\bibfnamefont {S.~Y.}\ \bibnamefont
  {Gavrilkin}}, \bibinfo {author} {\bibfnamefont {K.~V.}\ \bibnamefont
  {Mitsen}}, \bibinfo {author} {\bibfnamefont {N.~Y.}\ \bibnamefont
  {Shitsevalova}}, \bibinfo {author} {\bibfnamefont {A.~V.}\ \bibnamefont
  {Levchenko}}, \bibinfo {author} {\bibfnamefont {V.~B.}\ \bibnamefont
  {Filippov}}, \bibinfo {author} {\bibfnamefont {S.}~\bibnamefont {Gabani}},
  \bibinfo {author} {\bibfnamefont {K.}~\bibnamefont {Flachbart}}, \ and\
  \bibinfo {author} {\bibfnamefont {N.~E.}\ \bibnamefont {Sluchanko}},\
  }\href{https://doi.org/10.1134/s1063776113050014} {\bibfield  {journal}
  {\bibinfo  {journal} {J.~Exp. Theor. Phys.}\ }\textbf {\bibinfo {volume}
  {116}},\ \bibinfo {pages} {760} (\bibinfo {year} {2013})}\BibitemShut
  {NoStop}%
\bibitem [{\citenamefont {Matsumura}\ \emph
  {\textit{et~al.}}(2014{\natexlab{b}})\citenamefont {Matsumura}, \citenamefont
  {Michimura}, \citenamefont {Inami}, \citenamefont {Otsubo}, \citenamefont
  {Tanida}, \citenamefont {Iga},\ and\ \citenamefont
  {Sera}}]{MatsumuraMichimura14}%
  \BibitemOpen
  \bibfield  {author} {\bibinfo {author} {\bibfnamefont {T.}~\bibnamefont
  {Matsumura}}, \bibinfo {author} {\bibfnamefont {S.}~\bibnamefont
  {Michimura}}, \bibinfo {author} {\bibfnamefont {T.}~\bibnamefont {Inami}},
  \bibinfo {author} {\bibfnamefont {T.}~\bibnamefont {Otsubo}}, \bibinfo
  {author} {\bibfnamefont {H.}~\bibnamefont {Tanida}}, \bibinfo {author}
  {\bibfnamefont {F.}~\bibnamefont {Iga}}, \ and\ \bibinfo {author}
  {\bibfnamefont {M.}~\bibnamefont {Sera}},\
  }\href{https://link.aps.org/doi/10.1103/PhysRevB.89.014422} {\bibfield
  {journal} {\bibinfo  {journal} {Phys. Rev. B}\ }\textbf {\bibinfo {volume}
  {89}},\ \bibinfo {pages} {014422} (\bibinfo {year}
  {2014}{\natexlab{b}})}\BibitemShut {NoStop}%
\bibitem [{\citenamefont {Robinson}(2000)}]{Robinson00book}%
  \BibitemOpen
  \bibfield  {author} {\bibinfo {author} {\bibfnamefont {R.~A.}\ \bibnamefont
  {Robinson}},\ }in\ \href@noop {} {\emph {\bibinfo {booktitle} {Magnetism in
  Heavy Fermion Systems}}},\ \bibinfo {editor} {edited by\ \bibinfo {editor}
  {\bibfnamefont {H.~B.}\ \bibnamefont {Radousky}}}\ (\bibinfo  {publisher}
  {World Scientific, Singapore},\ \bibinfo {year} {2000})\ Chap.~\bibinfo
  {chapter} {4}, pp.\ \bibinfo {pages} {197--282}\BibitemShut {NoStop}%
\bibitem [{\citenamefont {Rossat-Mignod}\ \emph
  {\textit{et~al.}}(1988)\citenamefont {Rossat-Mignod}, \citenamefont
  {Regnault}, \citenamefont {Jacoud}, \citenamefont {Vettier}, \citenamefont
  {Lejay}, \citenamefont {Flouquet}, \citenamefont {Walker}, \citenamefont
  {Jaccard},\ and\ \citenamefont {Amato}}]{RossatMignod88}%
  \BibitemOpen
  \bibfield  {author} {\bibinfo {author} {\bibfnamefont {J.}~\bibnamefont
  {Rossat-Mignod}}, \bibinfo {author} {\bibfnamefont {L.~P.}\ \bibnamefont
  {Regnault}}, \bibinfo {author} {\bibfnamefont {J.~L.}\ \bibnamefont
  {Jacoud}}, \bibinfo {author} {\bibfnamefont {C.}~\bibnamefont {Vettier}},
  \bibinfo {author} {\bibfnamefont {P.}~\bibnamefont {Lejay}}, \bibinfo
  {author} {\bibfnamefont {J.}~\bibnamefont {Flouquet}}, \bibinfo {author}
  {\bibfnamefont {E.}~\bibnamefont {Walker}}, \bibinfo {author} {\bibfnamefont
  {D.}~\bibnamefont {Jaccard}}, \ and\ \bibinfo {author} {\bibfnamefont
  {A.}~\bibnamefont {Amato}},\
  }\href{https://doi.org/10.1016/0304-8853(88)90429-5} {\bibfield  {journal}
  {\bibinfo  {journal} {J.~Magn. Magn. Mater.}\ }\textbf {\bibinfo {volume}
  {76--77}},\ \bibinfo {pages} {376} (\bibinfo {year} {1988})}\BibitemShut
  {NoStop}%
\bibitem [{\citenamefont {Regnault}\ \emph {\textit{et~al.}}(1988)\citenamefont
  {Regnault}, \citenamefont {Erkelens}, \citenamefont {Rossat-Mignod},
  \citenamefont {Lejay},\ and\ \citenamefont {Flouquet}}]{RegnaultErkelens88}%
  \BibitemOpen
  \bibfield  {author} {\bibinfo {author} {\bibfnamefont {L.~P.}\ \bibnamefont
  {Regnault}}, \bibinfo {author} {\bibfnamefont {W.~A.~C.}\ \bibnamefont
  {Erkelens}}, \bibinfo {author} {\bibfnamefont {J.}~\bibnamefont
  {Rossat-Mignod}}, \bibinfo {author} {\bibfnamefont {P.}~\bibnamefont
  {Lejay}}, \ and\ \bibinfo {author} {\bibfnamefont {J.}~\bibnamefont
  {Flouquet}},\ }\href{https://link.aps.org/doi/10.1103/PhysRevB.38.4481}
  {\bibfield  {journal} {\bibinfo  {journal} {Phys. Rev.~B}\ }\textbf {\bibinfo
  {volume} {38}},\ \bibinfo {pages} {4481} (\bibinfo {year}
  {1988})}\BibitemShut {NoStop}%
\bibitem [{\citenamefont {Schr\"oder}\ \emph
  {\textit{et~al.}}(1998)\citenamefont {Schr\"oder}, \citenamefont {Aeppli},
  \citenamefont {Bucher}, \citenamefont {Ramazashvili},\ and\ \citenamefont
  {Coleman}}]{SchroederAeppli98}%
  \BibitemOpen
  \bibfield  {author} {\bibinfo {author} {\bibfnamefont {A.}~\bibnamefont
  {Schr\"oder}}, \bibinfo {author} {\bibfnamefont {G.}~\bibnamefont {Aeppli}},
  \bibinfo {author} {\bibfnamefont {E.}~\bibnamefont {Bucher}}, \bibinfo
  {author} {\bibfnamefont {R.}~\bibnamefont {Ramazashvili}}, \ and\ \bibinfo
  {author} {\bibfnamefont {P.}~\bibnamefont {Coleman}},\
  }\href{https://link.aps.org/doi/10.1103/PhysRevLett.80.5623} {\bibfield
  {journal} {\bibinfo  {journal} {Phys. Rev. Lett.}\ }\textbf {\bibinfo
  {volume} {80}},\ \bibinfo {pages} {5623} (\bibinfo {year}
  {1998})}\BibitemShut {NoStop}%
\bibitem [{\citenamefont {Stockert}\ \emph {\textit{et~al.}}(1998)\citenamefont
  {Stockert}, \citenamefont {L\"ohneysen}, \citenamefont {Rosch}, \citenamefont
  {Pyka},\ and\ \citenamefont {Loewenhaupt}}]{StockertLoehneysen98}%
  \BibitemOpen
  \bibfield  {author} {\bibinfo {author} {\bibfnamefont {O.}~\bibnamefont
  {Stockert}}, \bibinfo {author} {\bibfnamefont {H.~v.}\ \bibnamefont
  {L\"ohneysen}}, \bibinfo {author} {\bibfnamefont {A.}~\bibnamefont {Rosch}},
  \bibinfo {author} {\bibfnamefont {N.}~\bibnamefont {Pyka}}, \ and\ \bibinfo
  {author} {\bibfnamefont {M.}~\bibnamefont {Loewenhaupt}},\
  }\href{https://link.aps.org/doi/10.1103/PhysRevLett.80.5627} {\bibfield
  {journal} {\bibinfo  {journal} {Phys. Rev. Lett.}\ }\textbf {\bibinfo
  {volume} {80}},\ \bibinfo {pages} {5627} (\bibinfo {year}
  {1998})}\BibitemShut {NoStop}%
\bibitem [{\citenamefont {Stock}\ \emph {\textit{et~al.}}(2011)\citenamefont
  {Stock}, \citenamefont {Sokolov}, \citenamefont {Bourges}, \citenamefont
  {Tobash}, \citenamefont {Gofryk}, \citenamefont {Ronning}, \citenamefont
  {Bauer}, \citenamefont {Rule},\ and\ \citenamefont
  {Huxley}}]{StockSokolov11}%
  \BibitemOpen
  \bibfield  {author} {\bibinfo {author} {\bibfnamefont {C.}~\bibnamefont
  {Stock}}, \bibinfo {author} {\bibfnamefont {D.~A.}\ \bibnamefont {Sokolov}},
  \bibinfo {author} {\bibfnamefont {P.}~\bibnamefont {Bourges}}, \bibinfo
  {author} {\bibfnamefont {P.~H.}\ \bibnamefont {Tobash}}, \bibinfo {author}
  {\bibfnamefont {K.}~\bibnamefont {Gofryk}}, \bibinfo {author} {\bibfnamefont
  {F.}~\bibnamefont {Ronning}}, \bibinfo {author} {\bibfnamefont {E.~D.}\
  \bibnamefont {Bauer}}, \bibinfo {author} {\bibfnamefont {K.~C.}\ \bibnamefont
  {Rule}}, \ and\ \bibinfo {author} {\bibfnamefont {A.~D.}\ \bibnamefont
  {Huxley}},\ }\href{https://link.aps.org/doi/10.1103/PhysRevLett.107.187202}
  {\bibfield  {journal} {\bibinfo  {journal} {Phys. Rev. Lett.}\ }\textbf
  {\bibinfo {volume} {107}},\ \bibinfo {pages} {187202} (\bibinfo {year}
  {2011})}\BibitemShut {NoStop}%
\bibitem [{\citenamefont {Singh}\ \emph {\textit{et~al.}}(2011)\citenamefont
  {Singh}, \citenamefont {Thamizhavel}, \citenamefont {Lynn}, \citenamefont
  {Dhar}, \citenamefont {Rodriguez-Rivera},\ and\ \citenamefont
  {Herman}}]{SinghThamizhavel11}%
  \BibitemOpen
  \bibfield  {author} {\bibinfo {author} {\bibfnamefont {D.~K.}\ \bibnamefont
  {Singh}}, \bibinfo {author} {\bibfnamefont {A.}~\bibnamefont {Thamizhavel}},
  \bibinfo {author} {\bibfnamefont {J.~W.}\ \bibnamefont {Lynn}}, \bibinfo
  {author} {\bibfnamefont {S.}~\bibnamefont {Dhar}}, \bibinfo {author}
  {\bibfnamefont {J.}~\bibnamefont {Rodriguez-Rivera}}, \ and\ \bibinfo
  {author} {\bibfnamefont {T.}~\bibnamefont {Herman}},\
  }\href{https://doi.org/10.1038/srep00117} {\bibfield  {journal} {\bibinfo
  {journal} {Sci. Rep.}\ }\textbf {\bibinfo {volume} {1}},\ \bibinfo {pages}
  {117} (\bibinfo {year} {2011})}\BibitemShut {NoStop}%
\bibitem [{\citenamefont {Kimura}\ \emph {\textit{et~al.}}(2013)\citenamefont
  {Kimura}, \citenamefont {Nakatsuji}, \citenamefont {Wen}, \citenamefont
  {Broholm}, \citenamefont {Stone}, \citenamefont {Nishibori},\ and\
  \citenamefont {Sawa}}]{KimuraNakatsuji13}%
  \BibitemOpen
  \bibfield  {author} {\bibinfo {author} {\bibfnamefont {K.}~\bibnamefont
  {Kimura}}, \bibinfo {author} {\bibfnamefont {S.}~\bibnamefont {Nakatsuji}},
  \bibinfo {author} {\bibfnamefont {J.-J.}\ \bibnamefont {Wen}}, \bibinfo
  {author} {\bibfnamefont {C.}~\bibnamefont {Broholm}}, \bibinfo {author}
  {\bibfnamefont {M.~B.}\ \bibnamefont {Stone}}, \bibinfo {author}
  {\bibfnamefont {E.}~\bibnamefont {Nishibori}}, \ and\ \bibinfo {author}
  {\bibfnamefont {H.}~\bibnamefont {Sawa}},\
  }\href{https://doi.org/10.1038/ncomms2914} {\bibfield  {journal} {\bibinfo
  {journal} {Nature Commun.}\ }\textbf {\bibinfo {volume} {4}},\ \bibinfo
  {pages} {1934} (\bibinfo {year} {2013})}\BibitemShut {NoStop}%
\bibitem [{\citenamefont {Chan}\ and\ \citenamefont
  {Heine}(1973)}]{ChanHeine73}%
  \BibitemOpen
  \bibfield  {author} {\bibinfo {author} {\bibfnamefont {S.-K.}\ \bibnamefont
  {Chan}}\ and\ \bibinfo {author} {\bibfnamefont {V.}~\bibnamefont {Heine}},\
  }\href{http://stacks.iop.org/0305-4608/3/i=4/a=022} {\bibfield  {journal}
  {\bibinfo  {journal} {J.~Phys. F: Metal Physics}\ }\textbf {\bibinfo {volume}
  {3}},\ \bibinfo {pages} {795} (\bibinfo {year} {1973})}\BibitemShut {NoStop}%
\bibitem [{\citenamefont {Fawcett}(1988)}]{Fawcett88}%
  \BibitemOpen
  \bibfield  {author} {\bibinfo {author} {\bibfnamefont {E.}~\bibnamefont
  {Fawcett}},\ }\href{https://link.aps.org/doi/10.1103/RevModPhys.60.209}
  {\bibfield  {journal} {\bibinfo  {journal} {Rev. Mod. Phys.}\ }\textbf
  {\bibinfo {volume} {60}},\ \bibinfo {pages} {209} (\bibinfo {year}
  {1988})}\BibitemShut {NoStop}%
\bibitem [{\citenamefont {Borisenko}\ \emph
  {\textit{et~al.}}(2008)\citenamefont {Borisenko}, \citenamefont {Kordyuk},
  \citenamefont {Yaresko}, \citenamefont {Zabolotnyy}, \citenamefont {Inosov},
  \citenamefont {Schuster}, \citenamefont {B\"uchner}, \citenamefont {Weber},
  \citenamefont {Follath}, \citenamefont {Patthey},\ and\ \citenamefont
  {Berger}}]{BorisenkoKordyuk08}%
  \BibitemOpen
  \bibfield  {author} {\bibinfo {author} {\bibfnamefont {S.~V.}\ \bibnamefont
  {Borisenko}}, \bibinfo {author} {\bibfnamefont {A.~A.}\ \bibnamefont
  {Kordyuk}}, \bibinfo {author} {\bibfnamefont {A.~N.}\ \bibnamefont
  {Yaresko}}, \bibinfo {author} {\bibfnamefont {V.~B.}\ \bibnamefont
  {Zabolotnyy}}, \bibinfo {author} {\bibfnamefont {D.~S.}\ \bibnamefont
  {Inosov}}, \bibinfo {author} {\bibfnamefont {R.}~\bibnamefont {Schuster}},
  \bibinfo {author} {\bibfnamefont {B.}~\bibnamefont {B\"uchner}}, \bibinfo
  {author} {\bibfnamefont {R.}~\bibnamefont {Weber}}, \bibinfo {author}
  {\bibfnamefont {R.}~\bibnamefont {Follath}}, \bibinfo {author} {\bibfnamefont
  {L.}~\bibnamefont {Patthey}}, \ and\ \bibinfo {author} {\bibfnamefont
  {H.}~\bibnamefont {Berger}},\
  }\href{https://link.aps.org/doi/10.1103/PhysRevLett.100.196402} {\bibfield
  {journal} {\bibinfo  {journal} {Phys. Rev. Lett.}\ }\textbf {\bibinfo
  {volume} {100}},\ \bibinfo {pages} {196402} (\bibinfo {year}
  {2008})}\BibitemShut {NoStop}%
\bibitem [{\citenamefont {Ruderman}\ and\ \citenamefont
  {Kittel}(1954)}]{RudermanKittel54}%
  \BibitemOpen
  \bibfield  {author} {\bibinfo {author} {\bibfnamefont {M.~A.}\ \bibnamefont
  {Ruderman}}\ and\ \bibinfo {author} {\bibfnamefont {C.}~\bibnamefont
  {Kittel}},\ }\href{https://link.aps.org/doi/10.1103/PhysRev.96.99} {\bibfield
   {journal} {\bibinfo  {journal} {Phys. Rev.}\ }\textbf {\bibinfo {volume}
  {96}},\ \bibinfo {pages} {99} (\bibinfo {year} {1954})}\BibitemShut {NoStop}%
\bibitem [{\citenamefont {Kasuya}(1956)}]{Kasuya56}%
  \BibitemOpen
  \bibfield  {author} {\bibinfo {author} {\bibfnamefont {T.}~\bibnamefont
  {Kasuya}},\ }\href{https://doi.org/10.1143/PTP.16.45} {\bibfield  {journal}
  {\bibinfo  {journal} {Prog. Theor. Phys.}\ }\textbf {\bibinfo {volume}
  {16}},\ \bibinfo {pages} {45} (\bibinfo {year} {1956})}\BibitemShut {NoStop}%
\bibitem [{\citenamefont {Yafet}(1987)}]{Yafet87}%
  \BibitemOpen
  \bibfield  {author} {\bibinfo {author} {\bibfnamefont {Y.}~\bibnamefont
  {Yafet}},\ }\href{https://link.aps.org/doi/10.1103/PhysRevB.36.3948}
  {\bibfield  {journal} {\bibinfo  {journal} {Phys. Rev. B}\ }\textbf {\bibinfo
  {volume} {36}},\ \bibinfo {pages} {3948} (\bibinfo {year}
  {1987})}\BibitemShut {NoStop}%
\bibitem [{\citenamefont {Kim}\ \emph {\textit{et~al.}}(1996)\citenamefont
  {Kim}, \citenamefont {Lee},\ and\ \citenamefont {Lee}}]{KimLee96}%
  \BibitemOpen
  \bibfield  {author} {\bibinfo {author} {\bibfnamefont {J.~G.}\ \bibnamefont
  {Kim}}, \bibinfo {author} {\bibfnamefont {E.~K.}\ \bibnamefont {Lee}}, \ and\
  \bibinfo {author} {\bibfnamefont {S.}~\bibnamefont {Lee}},\
  }\href{https://link.aps.org/doi/10.1103/PhysRevB.54.6077} {\bibfield
  {journal} {\bibinfo  {journal} {Phys. Rev. B}\ }\textbf {\bibinfo {volume}
  {54}},\ \bibinfo {pages} {6077} (\bibinfo {year} {1996})}\BibitemShut
  {NoStop}%
\bibitem [{\citenamefont {Aristov}(1997)}]{Aristov97}%
  \BibitemOpen
  \bibfield  {author} {\bibinfo {author} {\bibfnamefont {D.~N.}\ \bibnamefont
  {Aristov}},\ }\href{https://link.aps.org/doi/10.1103/PhysRevB.55.8064}
  {\bibfield  {journal} {\bibinfo  {journal} {Phys. Rev. B}\ }\textbf {\bibinfo
  {volume} {55}},\ \bibinfo {pages} {8064} (\bibinfo {year}
  {1997})}\BibitemShut {NoStop}%
\bibitem [{\citenamefont {Litvinov}\ and\ \citenamefont
  {Dugaev}(1998)}]{LitvinovDugaev98}%
  \BibitemOpen
  \bibfield  {author} {\bibinfo {author} {\bibfnamefont {V.~I.}\ \bibnamefont
  {Litvinov}}\ and\ \bibinfo {author} {\bibfnamefont {V.~K.}\ \bibnamefont
  {Dugaev}},\ }\href{https://link.aps.org/doi/10.1103/PhysRevB.58.3584}
  {\bibfield  {journal} {\bibinfo  {journal} {Phys. Rev. B}\ }\textbf {\bibinfo
  {volume} {58}},\ \bibinfo {pages} {3584} (\bibinfo {year}
  {1998})}\BibitemShut {NoStop}%
\bibitem [{\citenamefont {Brown}\ \emph {\textit{et~al.}}(1985)\citenamefont
  {Brown}, \citenamefont {Caudron}, \citenamefont {Fert}, \citenamefont
  {Givord},\ and\ \citenamefont {Pureur}}]{BrownCaudron85}%
  \BibitemOpen
  \bibfield  {author} {\bibinfo {author} {\bibfnamefont {P.~J.}\ \bibnamefont
  {Brown}}, \bibinfo {author} {\bibfnamefont {R.}~\bibnamefont {Caudron}},
  \bibinfo {author} {\bibfnamefont {A.}~\bibnamefont {Fert}}, \bibinfo {author}
  {\bibfnamefont {D.}~\bibnamefont {Givord}}, \ and\ \bibinfo {author}
  {\bibfnamefont {P.}~\bibnamefont {Pureur}},\
  }\href{\doibase/10.1051/jphyslet:0198500460230113900} {\bibfield  {journal}
  {\bibinfo  {journal} {J.~Physique Lett.}\ }\textbf {\bibinfo {volume} {46}},\
  \bibinfo {pages} {1139} (\bibinfo {year} {1985})}\BibitemShut {NoStop}%
\bibitem [{\citenamefont {Fretwell}\ \emph {\textit{et~al.}}(1999)\citenamefont
  {Fretwell}, \citenamefont {Dugdale}, \citenamefont {Alam}, \citenamefont
  {Hedley}, \citenamefont {Rodriguez-Gonzalez},\ and\ \citenamefont
  {Palmer}}]{FretwellDugale99}%
  \BibitemOpen
  \bibfield  {author} {\bibinfo {author} {\bibfnamefont {H.~M.}\ \bibnamefont
  {Fretwell}}, \bibinfo {author} {\bibfnamefont {S.~B.}\ \bibnamefont
  {Dugdale}}, \bibinfo {author} {\bibfnamefont {M.~A.}\ \bibnamefont {Alam}},
  \bibinfo {author} {\bibfnamefont {D.~C.~R.}\ \bibnamefont {Hedley}}, \bibinfo
  {author} {\bibfnamefont {A.}~\bibnamefont {Rodriguez-Gonzalez}}, \ and\
  \bibinfo {author} {\bibfnamefont {S.~B.}\ \bibnamefont {Palmer}},\
  }\href{\doibase/10.1103/PhysRevLett.82.3867} {\bibfield  {journal} {\bibinfo
  {journal} {Phys. Rev. Lett.}\ }\textbf {\bibinfo {volume} {82}},\ \bibinfo
  {pages} {3867} (\bibinfo {year} {1999})}\BibitemShut {NoStop}%
\bibitem [{\citenamefont {Rau}\ \emph {\textit{et~al.}}(2014)\citenamefont
  {Rau}, \citenamefont {Lee},\ and\ \citenamefont {Kee}}]{RauLee14}%
  \BibitemOpen
  \bibfield  {author} {\bibinfo {author} {\bibfnamefont {J.~G.}\ \bibnamefont
  {Rau}}, \bibinfo {author} {\bibfnamefont {E.~K.-H.}\ \bibnamefont {Lee}}, \
  and\ \bibinfo {author} {\bibfnamefont {H.-Y.}\ \bibnamefont {Kee}},\
  }\href{https://link.aps.org/doi/10.1103/PhysRevLett.112.077204} {\bibfield
  {journal} {\bibinfo  {journal} {Phys. Rev. Lett.}\ }\textbf {\bibinfo
  {volume} {112}},\ \bibinfo {pages} {077204} (\bibinfo {year}
  {2014})}\BibitemShut {NoStop}%
\bibitem [{\citenamefont {Iqbal}\ \emph {\textit{et~al.}}(2015)\citenamefont
  {Iqbal}, \citenamefont {Jeschke}, \citenamefont {Reuther}, \citenamefont
  {Valent\'{\i}}, \citenamefont {Mazin}, \citenamefont {Greiter},\ and\
  \citenamefont {Thomale}}]{IqbalJeschke15}%
  \BibitemOpen
  \bibfield  {author} {\bibinfo {author} {\bibfnamefont {Y.}~\bibnamefont
  {Iqbal}}, \bibinfo {author} {\bibfnamefont {H.~O.}\ \bibnamefont {Jeschke}},
  \bibinfo {author} {\bibfnamefont {J.}~\bibnamefont {Reuther}}, \bibinfo
  {author} {\bibfnamefont {R.}~\bibnamefont {Valent\'{\i}}}, \bibinfo {author}
  {\bibfnamefont {I.~I.}\ \bibnamefont {Mazin}}, \bibinfo {author}
  {\bibfnamefont {M.}~\bibnamefont {Greiter}}, \ and\ \bibinfo {author}
  {\bibfnamefont {R.}~\bibnamefont {Thomale}},\
  }\href{https://link.aps.org/doi/10.1103/PhysRevB.92.220404} {\bibfield
  {journal} {\bibinfo  {journal} {Phys. Rev.~B}\ }\textbf {\bibinfo {volume}
  {92}},\ \bibinfo {pages} {220404} (\bibinfo {year} {2015})}\BibitemShut
  {NoStop}%
\bibitem [{\citenamefont {Bieri}\ \emph {\textit{et~al.}}(2016)\citenamefont
  {Bieri}, \citenamefont {Lhuillier},\ and\ \citenamefont
  {Messio}}]{BierliLhuillier16}%
  \BibitemOpen
  \bibfield  {author} {\bibinfo {author} {\bibfnamefont {S.}~\bibnamefont
  {Bieri}}, \bibinfo {author} {\bibfnamefont {C.}~\bibnamefont {Lhuillier}}, \
  and\ \bibinfo {author} {\bibfnamefont {L.}~\bibnamefont {Messio}},\
  }\href{https://link.aps.org/doi/10.1103/PhysRevB.93.094437} {\bibfield
  {journal} {\bibinfo  {journal} {Phys. Rev.~B}\ }\textbf {\bibinfo {volume}
  {93}},\ \bibinfo {pages} {094437} (\bibinfo {year} {2016})}\BibitemShut
  {NoStop}%
\bibitem [{\citenamefont {Tymoshenko}\ \emph
  {\textit{et~al.}}(2017)\citenamefont {Tymoshenko}, \citenamefont
  {Onykiienko}, \citenamefont {M\"uller}, \citenamefont {Thomale},
  \citenamefont {Rachel}, \citenamefont {Cameron}, \citenamefont
  {Portnichenko}, \citenamefont {Efremov}, \citenamefont {Tsurkan},
  \citenamefont {Abernathy}, \citenamefont {Ollivier}, \citenamefont
  {Schneidewind}, \citenamefont {Piovano}, \citenamefont {Felea}, \citenamefont
  {Loidl},\ and\ \citenamefont {Inosov}}]{TymoshenkoOnykiienko17}%
  \BibitemOpen
  \bibfield  {author} {\bibinfo {author} {\bibfnamefont {Y.~V.}\ \bibnamefont
  {Tymoshenko}}, \bibinfo {author} {\bibfnamefont {Y.~A.}\ \bibnamefont
  {Onykiienko}}, \bibinfo {author} {\bibfnamefont {T.}~\bibnamefont
  {M\"uller}}, \bibinfo {author} {\bibfnamefont {R.}~\bibnamefont {Thomale}},
  \bibinfo {author} {\bibfnamefont {S.}~\bibnamefont {Rachel}}, \bibinfo
  {author} {\bibfnamefont {A.~S.}\ \bibnamefont {Cameron}}, \bibinfo {author}
  {\bibfnamefont {P.~Y.}\ \bibnamefont {Portnichenko}}, \bibinfo {author}
  {\bibfnamefont {D.~V.}\ \bibnamefont {Efremov}}, \bibinfo {author}
  {\bibfnamefont {V.}~\bibnamefont {Tsurkan}}, \bibinfo {author} {\bibfnamefont
  {D.~L.}\ \bibnamefont {Abernathy}}, \bibinfo {author} {\bibfnamefont
  {J.}~\bibnamefont {Ollivier}}, \bibinfo {author} {\bibfnamefont
  {A.}~\bibnamefont {Schneidewind}}, \bibinfo {author} {\bibfnamefont
  {A.}~\bibnamefont {Piovano}}, \bibinfo {author} {\bibfnamefont
  {V.}~\bibnamefont {Felea}}, \bibinfo {author} {\bibfnamefont
  {A.}~\bibnamefont {Loidl}}, \ and\ \bibinfo {author} {\bibfnamefont {D.~S.}\
  \bibnamefont {Inosov}},\
  }\href{https://link.aps.org/doi/10.1103/PhysRevX.7.041049} {\bibfield
  {journal} {\bibinfo  {journal} {Phys. Rev.~X}\ }\textbf {\bibinfo {volume}
  {7}},\ \bibinfo {pages} {041049} (\bibinfo {year} {2017})}\BibitemShut
  {NoStop}%
\bibitem [{\citenamefont {Schmidt}\ and\ \citenamefont
  {Thalmeier}(2017)}]{SchmidtThalmeier17}%
  \BibitemOpen
  \bibfield  {author} {\bibinfo {author} {\bibfnamefont {B.}~\bibnamefont
  {Schmidt}}\ and\ \bibinfo {author} {\bibfnamefont {P.}~\bibnamefont
  {Thalmeier}},\ }\href{https://doi.org/10.1016/j.physrep.2017.06.004}
  {\bibfield  {journal} {\bibinfo  {journal} {Phys. Rep.}\ }\textbf {\bibinfo
  {volume} {703}},\ \bibinfo {pages} {1} (\bibinfo {year} {2017})}\BibitemShut
  {NoStop}%
\end{thebibliography}%
\vspace{-4pt}

\end{document}